\documentclass[onecolumn]{mn2e}
\newif\ifAMStwofonts

\def\pmb#1{\mbox{\boldmath$#1$}}
\def\gtsim {>\kern-1.2em\lower1.1ex\hbox{$\sim$}}
\def\ltsim {<\kern-1.2em\lower1.1ex\hbox{$\sim$}}
\def\gtsim {>\kern-1.2em\lower1.1ex\hbox{$\sim$}}
\def\ltsim {<\kern-1.2em\lower1.1ex\hbox{$\sim$}}
\def\ref{\hangindent=1pc \hangafter=1 \noindent}

\def\cvp#1{\left\{#1\right\}}
\def\sdp#1{\left[#1\right]}
\def\kgp#1{\left<#1\right>}
\def\unitm{\pmb{1}}

\def\be{\begin{equation}}
\def\ee{\end{equation}}

\usepackage{epsfig}
\begin{document}



\title{Axisymmetric oscillations of magnetic neutron stars}
\author[U. Lee]{Umin Lee$^1$\thanks{E-mail: lee@astr.tohoku.ac.jp}
\\$^1$Astronomical Institute, Tohoku University, Sendai, Miyagi 980-8578, Japan}

\date{Typeset \today ; Received / Accepted}
\maketitle


\begin{abstract} 
We calculate axisymmetric oscillations of rotating neutron stars composed  of 
the surface fluid ocean, solid crust, and fluid core, taking account of a dipole 
magnetic field as strong as $B_S\sim 10^{15}$G at the surface.
The adiabatic oscillation equations for the solid crust threaded by a dipole magnetic field
are derived in Newtonian dynamics, on the assumption that 
the axis of rotation is aligned 
with the magnetic axis so that perturbations on the equilibrium can be represented 
by series expansions
in terms of spherical harmonic functions $Y_l^m(\theta,\phi)$ 
with different degrees $l$ for a given azimuthal wave number $m$ around the the magnetic axis.
Although the three component models can support a rich variety of oscillation modes,
axisymmetric ($m=0$)
toroidal $_{l}t_n$ and spheroidal $_ls_n$ shear waves propagating in the solid crust 
are our main concerns, where $l$ and $n$ denote the harmonic degree and the radial order of 
the modes, respectively.
In the absence of rotation, axisymmetric spheroidal and toroidal modes
are completely decoupled, and we consider the effects of rotation on the oscillation modes
only in the limit of slow rotation.
We find that the oscillation frequencies of the fundamental toroidal torsional modes 
$_{l}t_n$
in the crust are hardly affected by the magnetic field as strong as $B_S\sim 10^{15}$G 
at the surface.
As the radial order $n$ of the shear modes in the crust becomes higher, 
however, both spheroidal and toroidal modes
become susceptible to the magnetic field and their frequencies in general get higher with
increasing $B_S$.
We also find that the surface $g$ modes and the crust/ocean interfacial modes
are suppressed by a strong magnetic field, and that
there appear magnetic modes in the presence of a strong magnetic field.
\end{abstract}


\section{Introduction}

Recent discoveries of quasi-periodic oscillations (QPOs) in the giant flares 
of Soft Gamma-Ray Repeaters
SGR 1806-20 (Israel et al 2005) and SGR 1900+14 (Strohmayer \& Watts 2005) and
the confirmation of the discoveries (Watts \& Strohmayer 2006, Strohmayer \& Watts 2006)
have made promising asteroseismology for magnetors,
neutron stars with an extremely strong magnetic field
(see, e.g., Woods \& Thompson 2004 for a review on SGRs).
Israel et al (2005) discovered QPOs of frequencies $\sim18$, $\sim 30$ and $\sim$90Hz 
in the tail of the SGR 1806-20 hyperflare observed December 2004, and suggested 
that the 30Hz and 90Hz QPOs
could be caused by seismic vibrations of the neutron star crust (see, e.g., Duncan 1998).
Later on, in the hyperflare of SGR 1900+14, Strohmayer \& Watts (2005) found QPOs 
of frequencies 28, 53.5, 84, and 155 Hz, and claimed that the QPOs
could be identified as low $l$ fundamental toroidal torsional modes 
in the solid crust of the neutron star.
Although the interpretation in terms of the crustal torsional modes is promising,
the mode identification cannot always be definite.
In fact, if different classes of oscillation modes can generate similar
periodicities and information other than the periods are not
available for the QPOs,
it is usually difficult to assign a class of oscillation modes to the periods observed
in preference to other classes of modes. 
This is particularly true for high frequency QPOs 
(e.g., 625Hz QPO, Watts \& Strohmayer 2006; 1835Hz QPO and less significant QPOs at
720 and 2384 Hz in SGR 1806-20, Strohmayer \& Watts 2006), 
for which there exist various classes of modes that can 
generate the periodicities observed.
In this case, the pattern of observed frequencies, that is, the observed frequency spectrum
could be a key for mode identification and hence understanding the underlying neutron stars themselves.
McDermott et al (1988) have carried out detailed modal analyses for neutron star models with
a solid crust, but without the effects of magnetic field and rotation
on the oscillation modes.
Since the detected QPOs are thought to come from magnetors inferred to possess a strong magnetic field,
it is useful to calculate oscillations of rotating neutron stars that have 
a solid crust and a strong magnetic field.

An extensive theoretical modal analysis of magnetic neutron stars having a solid crust
was first carried out by Carroll et al (1986), who however employed a cylindrical geometry for
the analysis of the neutron star models, 
and assumed a uniform magnetic field whose axis is parallel to
the axis of the cylinder. 
The maximum strength of the magnetic field they examined is $B_S\sim 10^{12}$G,
which may be too weak for mangetors for which $B_S\gtsim 10^{14}$G is inferred.
Carroll et al (1986) have suggested
the existence of Alfv\'en modes and also found the transformation of the $g$ modes
in the surface ocean into magnetic modes as the field becomes stronger.
More recently, Piro (2005) solved a simplified set of oscillation equations for
toroidal torsional modes employing improved shear modulus (Strohmayer et al 1991) and
up to date microphysics for equations of state to construct back ground neutron star models.
Since the fluid core was ignored in
the calculations by Carroll et al (1986) and Piro (2005), 
no reliable modal analyses were possible for spheroidal modes that can have
substantial amplitudes in the core.
In this paper, we calculate various oscillation modes of neutron star models that have
a solid crust and are threaded by a dipole magnetic field.
Since a non-radial mode of a spherical neutron star thereaded by a dipole magnetic field
cannot be represented by a single spherical harmonic function, we employ series
expansions to represent the perturbations accompanied by a mode.
In \S2, we give a brief description of method of solution we empoy, numerical results
are given in \S 3, and
\S 4 is for conclusions.
The oscillation equations are given in Appendix as well as boundary and jump conditions
used in this paper.

\section{Method of solution}

\subsection{Perturbation Equations}

To derive oscillation equations for a magnetized and rotating neutron star with a solid crust, 
we follow Carroll et al (1986), Lee \& Strohmayer (1996), and Lee (2004).
We consider no general relativistic effects on oscillations, that is, the oscillation
equations are derived in Newtonian dynamics.
We employ spherical polar
coordinates $(r,\theta,\phi)$, whose origin is at the center of the star and the axis of rotation
is the axis of $\theta=0$.
We assume a dipole magnetic field given by
\begin{equation}
\pmb{B}=\mu_m\nabla(\cos\theta/r^2),
\end{equation}
where $\mu_m$ is the magnetic dipole moment.
For simplicity, we also assume that the magnetic axis coincides with the rotation axis.
Since the dipole field is a force-free field such that $(\nabla\times \pmb{B})\times\pmb{B}=0$, 
the field does not influence the equilibrium structure of the star.
Assuming the axis of rotation is also the magnetic axis,
the temporal and angular dependence of perturbations can be given by a single factor $e^{i(m\phi+\omega t)}$,
where $m$ is the azimuthal wavenumber around the rotation axis and $\omega\equiv\sigma+m\Omega$ 
is the oscillation frequency observed in the corotating frame of the star where
$\sigma$ is the oscillations frequency in an inertial frame, and $\Omega$ is the angular frequency
of rotation.
The linearized basic equations applied in the solid crustal region of the star are then given by
\be
-\omega^2\pmb{\xi}+2i\omega\pmb{\Omega}\times\pmb{\xi}=
{1\over\rho}\nabla\cdot\pmb{\sigma}^\prime
-{\rho^\prime\over\rho^2}\nabla\cdot\pmb{\sigma}
+{1\over4\pi\rho}\left(\nabla\times\pmb{B}^\prime\right)\times\pmb{B}
-{1\over 4\pi c}i\omega\pmb{E}^\prime\times\pmb{B},
\ee
\be
\rho^\prime+\nabla\cdot(\rho\pmb{\xi})=0,
\ee
\be
{\rho^\prime\over\rho}={1\over\Gamma_1}{p^\prime\over p}-\xi_rA,
\ee
\be
\pmb{B}^\prime=\nabla\times(\pmb{\xi}\times\pmb{B}),
\ee
\be
\pmb{E}^\prime=-i\omega\pmb{\xi}\times\pmb{B}/c,
\ee
where $\rho$ is the mass density, $p$ is the pressure, 
$c$ is the velocity of light, $\pmb{\xi}$ is the displacement vector,
$\pmb{B}^\prime$ and $\pmb{E}^\prime$ are the Euler perturbations of 
magnetic and electric fields, respectively,
and the other physical quantities with a prime $(^\prime)$ denote their Euler perturbations,
and $A$ is the Schwartzshild discriminant defined by
\be
A={d\ln\rho\over dr}-{1\over\Gamma_1}{d\ln p\over dr},
\ee
and
\be
\Gamma_1=\left({\partial\ln p\over\partial\ln\rho}\right)_{ad}.
\ee
Note that we have employed the Cowling approximation neglecting 
the Eulerian perturbation of the gravitational potential, and that no effects of rotational
deformation are included.
In equation (2), $\pmb{\sigma}^\prime$ denotes the Euler perturbation of the stress tensor and is
obtained from the Lagrangian perturbation defined in Cartesian coordinates by
\be
\delta\sigma_{ij}=(\Gamma_1pu)\delta_{ij}+2\mu (u_{ij}-{1\over 3}u\delta_{ij})
\ee
with $u_{ij}$ being the strain tensor defined by
\be
u_{ij}={1\over 2}\left({\partial\xi_i\over\partial x_j}+{\partial\xi_j\over\partial x_i}\right),
\ee
where $\delta_{ij}$ denotes Kronecker delta, $\mu$ is the shear modulus, and 
$u=\sum_{l=1}^3u_{ll}$.
The last term on the right-hand side of equation (2) represents the contribution
from the displacement current, where infinite conductivity has been assumed.
The linearized basic equations for a fluid region
may be obtained by simply replacing the terms $\nabla\cdot\pmb{\sigma}$ and $\nabla\cdot\pmb{\sigma}^\prime$
by $-\nabla p$ and $-\nabla p^\prime$, respectively.

Since the angular dependence of perturbations on a rotating and magnetized star cannot be represented by
a single spherical harmonic function, we expand the perturbed quantities in terms of 
spherical harmonic functions $Y_l^m$ with different $l$s for a given $m$, on the assumption that
the axis of rotation coincides with that of the magnetic field.
The displacement vector $\pmb{\xi}$ and the perturbed magnetic field $\pmb{B}^\prime$
are then approximately represented by finite series expansions of length $j_{\rm max}$ as
\be
{\pmb{\xi}\over r}=\sum_{j=1}^{j_{\rm max}}\left\{\left[S_{l_j}(r)+H_{l_j}(r)\nabla \right]Y^m_{l_j}(\theta,\phi)
+T_{l^\prime_j}(r)~\pmb{e}_r\times\nabla Y^m_{l^\prime_j}(\theta,\phi)\right\}e^{i\omega t},
\ee
and 
\be
{\pmb{B}^\prime\over B_0(r)}=\sum_{j=1}^{j_{\rm max}}\left\{\left[b^S_{l^\prime_j}(r)
+b^H_{l^\prime_j}(r)\nabla \right]Y^m_{l^\prime_j}(\theta,\phi)
+b^T_{l_j}(r)~\pmb{e}_r\times\nabla Y^m_{l_j}(\theta,\phi)\right\}e^{i\omega t},
\ee
and the pressure perturbation, $p^\prime$, for example, is given by
\be
p^\prime=\sum_{j=1}^{j_{\rm max}}p^\prime_{l_j}(r)Y_{l_j}^m(\theta,\phi)e^{i\omega t}.
\ee
where $B_0(r)=\mu_m/r^3$,
and $l_j=|m|+2(j-1)$ and $l^\prime_j=l_j+1$ for even modes, and 
$l_j=|m|+2j-1$ and $l^\prime_j=l_j-1$ for odd modes, respectively, and $j=1,~2,~3,~\cdots, ~j_{\rm max}$.
In this paper, we have used $j_{\rm max}=12$.
Substituting the expansions (11)$\sim$(13) into the linearized basic equations (2)$\sim$(6), 
and making use of equations (9) and (10), 
we obtain a finite set of coupled linear ordinary differential equations for the expansion coefficients
such as $S_{l_j}(r)$ and $b^S_{l^\prime_j}(r)$, which we call
the oscillation equations solved in the solid crust.
The oscillation equations to be solved in magnetic fluid regions can be derived 
in the same manner (e.g., Lee 2004).
The sets of oscillation equations thus obtained for the solid and fluid regions
are given in Appendix A for the case of axisymmetric modes with $m=0$.
The boundary conditions at the center and the surface of the star and the jump conditions
imposed at the interfaces between fluid and crustal regions are discussed in Appendix B.
It is important to note that in the case of $\Omega=0$
the oscillation equations for axisymmetric modes with $m=0$
are decoupled into those for spheroidal modes and toroidal modes,
respectively, and that axisymmetric spheroidal and toroidal modes are coupled
only through the effects of rotation.
For non-axisymmetric modes with $m\not=0$, however
there occurs no decoupling between spheroidal modes and
toroidal modes even for $\Omega=0$.

\subsection{Non-Magnetic Core}

In a fluid region where the magnetic pressure $p_B\equiv B_0^2(r)/8\pi$ is much smaller than
the gas pressure $p$, that is, 
where the Alfv\'en velocity $v_A=\sqrt{2p_B/\rho}$ is much smaller
than the sound velocity $\sim\sqrt{p/\rho}$ or $r\omega$, 
the magnetic perturbations $\pmb{b}^H$ and $i\pmb{b}^T$ suffer
very rapid spatial oscillations except for extremely low frequency oscillations
(e.g., Lee 2004, see also Biront et al 1982, Roberts \& Soward 1983, Cambell \& Papaloizou 1986).
This rapid spatial oscillation happens in the fluid core 
for both spheroidal and toroidal modes.
For axisymmetric toroidal modes, for example,
the dimensionless wavenumber $k_r$ in the radial direction may be given by
\be
k_r\sim r\omega/v_A,
\ee
and the wavenumber can be very large
in the magnetic fluid region immediately below the crust where $p_B/p\ll 1$
even for $B_S=B_0(R)\sim10^{15}$G at the surface.
Since this kind of rapid spatial oscillations usually do not occur in the solid crust,
there exists an abrupt change in the property of the magnetic eigenfunctions
across the boundary between the solid crust and the fluid core.
Numerically, this means that we need an extremely large number of mesh points
to correctly calculate very short magnetic perturbations in the fluid core, 
which is not feasible.
Since the rapid spatial oscillations in the magnetic perturbations come from
the fact that the magnetic pressure significantly weaker compared to
the gas pressure in the fluid core, it is legitimate to assume that the fluid core 
is non-magnetic so that
the oscillation equations for non-magnetic fluid stars (e.g., Lee \& Saio 1990) could be used.
Assuming non-magnetic core, we neglect a possible dissipation
of the oscillation energy due to Joule heating in the core.

\section{Numerical Results}

In this paper we mainly discuss the case of $\Omega=0$, in which
axisymmetric $(m=0)$ toroidal and spheroidal modes are decoupled.
We briefly discuss the effects of rotation on the axisymmetric modes,
assuming slow rotation so that
$\left|\Omega/\sqrt{GM/R^3}\right|\ll 1$, in which case toroidal and spheroidal modes
are only weakly coupled.

\subsection{Equilibrium Models}

The neutron star models we use consist of the surface fluid ocean, the solid crust, and 
the fluid core and  
are the same as those employed by McDermott et al (1988) for
modal calculations without rotation and magnetic fields.
The models are obtained from the fully general relativistic evolutionary cooling calculations 
of neutron stars by Richardson et al (1980).
The outer crust extends down to the neutron drip point at $\rho=4.3\times 10^{11}$ g cm$^{-3}$ 
and is
assumed to consist of bare Fe nuclei embedded in a uniform, neutralizing, degenerate electron gas.
The strength of the Coulomb interaction between ions is characterized by the dimensionless parameter
$\Gamma$, the ratio of the Coulomb energy to the termal energy $k_BT$ 
with $k_B$ being Boltzmann constant.
The matter of the outer crust is assumed to undergo a first order fluid/solid phase transition
at $\Gamma=155$, and there exists the fluid ocean above the crystallization boundary at $\Gamma=155$.
In addition to the thermodynamic contributions from the nuclear and electronic kinetic energies, 
the equation of state used in the outer crust includes contributions from photons, 
the Coulomb interactions between nuclei and electrons, 
and nuclear vibrations and rotations.
The inner crust extends from the neutron drip point at $4.3\times 10^{11}$ g cm$^{-3}$
to the base of the crust at $2.4\times 10^{14}$ g cm$^{-3}$ and it is assumed to consist of
nuclei with $Z\sim40$, degenerate electrons, and degenerate, nonrelativistic neutrons, where
the actual composition of the nuclei in the inner crust is taken from Negele \& Vautherin (1973).
For the inner crust, the zero temperature equation of state is that by Negele \& Vautherin (1973),
and the leading order thermal corrections for the nuclei, electrons, and free neutrons are also included.
At densities greater than $2.4\times 10^{14}$ g cm$^{-3}$, the lattice is assumed to dissolve, and 
the core
of the neutron star is taken to consist of a mixture of free and highly degenerate neutrons,
protons, and electrons.
The equation of state in the fluid core is that by Baym, Bethe, \& Pethick (1971), and
are added the leading thermal corrections for the three species.
The outer most fluid envelope models are from the calculations by Gudmundsson et al (1983),
and are fitted to the evolutionaly core models by Richrdson et al (1980) at
mass density $\rho \approx 10^{10}$g cm$^{-3}$.
The more details of the models for oscillation calculations
can be found in McDermott, van Horn, \& Hansen (1988).

The shear modulus $\mu$ of the solid lattice is that given by Pandharipande, Pines, \& Smith (1976):
\be
\mu=0.3711{Z^2e^2n_N^{4/3}\over 2^{1/3}},
\ee
where $n_N$ is the number density of the nuclei.
An improved calculation for the shear modulus was given by Strohmayer et al (1991).
This imporved shear modulus was used by several authors to calculate crustal oscillations
of neutron stars, for example, by Duncun (1998) and Piro (2005).
In this paper, we did not attempt to use imporved models for modal calculations
since our main concerns are to investigate the modal properties of magnetic
neutron stars and not to try a detailed comparison between theoretical predictions
and observational results for the QPOs from the SGRs.

For mode calculations in this paper, we mainly use neutron star models named NS05T7
and NS13T8, and 
the mass and radius of the former are $M=0.503M_\odot$ and $R=9.839$km, respectively, and
those of the latter are $M=1.326M_\odot$ and $R=7.853$km, respectively.
Both of the models have a solid crust, and the thickness of the crust is
$\Delta r/R\sim0.24$ for NS05T7 and $\Delta r/R\sim 0.055$ for NS13T8.
The thickness of the surface ocean, on the other hand, 
is $\Delta r/R\sim3.8\times10^{-5}$ for NS05T7 and $\Delta r/R\sim 2.3\times10^{-3}$ for NS13T8,
respectively.
The details of the physical parameters of these models are given in McDermott et al (1988).

\subsection{Toroidal Modes}

In Figure 1, the normalized frequencies $\bar\omega\equiv\omega/\sqrt{GM/R^3}$ of the 
axisymmetric ($m=0$) toroidal torsional modes $_{l^\prime}t_{n}$ are plotted versus $B_S$
for the models NS05T7 (top panel) and NS13T8 (bottom panel), where
$B_S\equiv\mu_m/R^3$ is the strength of the dipole magnetic field at the surface.
The frequencies $\bar\omega$ are local ones and for convenience the redshifted frequencies 
defined by $\nu_\infty\equiv \omega(1-2GM/Rc^2)^{1/2}/2\pi$
are also displayed on the right axis of the panels.
We find that the frequencies $\bar\omega$ of 
the fundamental toroidal modes are only weakly affected by the dipole magnetic field
of strength as large as $B_S\sim10^{15}$G at the surface, which is consistent with
the results by Piro (2005).
Note that the frequencies of the fundamental modes $_{l^\prime}t_0$ show 
a slight decrease with increasing $B_S$ for low values of $l^\prime$.
The overtone torsional modes $_{l^\prime}t_n$ having $n\ge 1$ become more susceptible to the magnetic field
as the radial order $n$ gets higher and the wavelengths in the radial direction
become shorter so that several wavelengths of the modes are spaned by the strong magnetic region 
in the outer most envelope (or crust). We find as a general trend the frequency of the overtone modes
increases with increasing $B_S$. Besides the magnetic effect just mentioned, there exists
a different kind of magnetic effects closely related to the fact that the separation of variables
using a single spherical harmonic function is not possible for the oscillations of stars with a dipole magnetic field.
In this case we may consider
crustal toroidal modes with different $l^\prime$s but with the same parity are coupled 
in the presence of the magnetic field.
Since the toroidal modes with a given radial order $n\ge 1$
are nearly degenerate in frequency for different values of $l^\prime$, the coupling effect brings about
interference between them, which becomes significant as the field strength increases.
%
As discussed by McDermott et al (1988) for non-magnetized neutron stars, 
the frequencies of the fundamental toroidal modes
with different $l^\prime$s approximately scale as 
\footnote{As pointed out by one of the referees, the frequency scaling formula given by equation (16) is correct only
for $l^\prime\gg1$, and a much better scaling formula should be
$\bar\omega_{l^\prime}\approx\sqrt{(l^\prime-1)(l^\prime+2)/4}~\bar\omega_{l^\prime=2}$ as one
can easily verify by solving the simple case of a uniform density, uniform shear modulus oscillating crust.}
\be
\bar\omega_{l^\prime}\approx\sqrt{l^\prime(l^\prime+1)/6}~\bar\omega_{l^\prime=2},
\ee
which is also confirmed even in the presence of a strong magnetic field $B_S\sim 10^{15}$G.

The eigenfunctions $iT_{l^\prime}$ and $ib^T_l$ 
of the axisymmetric fundamental $_{2}t_{0}$ mode
of the model NS05T7
are depicted in Figure 2 for $B_S=10^{12}$G in panels (a) and (b)
and for $B_S=10^{15}$G in panels (c) and (d),
where the solid and dashed lines stand for $l^\prime=2$ and $4$ for $iT_{l^\prime}$ and
for $l=1$ and $3$ for $ib^T_l$,
and the amplitude normalization is given by $iT_{l^\prime=2}=1$ at the surface of the star.
Different from the case with no magnetic field, 
the toroidal components $iT_{l^\prime}$ of the displacement vector
is continuous at the crust/ocean interface (see Carroll et al 1986).
However, since we have assumed the fluid core is non-magnetic, no toroidal components
$iT_{l^\prime}$ and $ib^T_l$ exist in the core for $\Omega=0$.
Both for $B_S=10^{12}$G and $B_S=10^{15}$G, $iT_{l^\prime=2}$ dominates 
$iT_{l^\prime=4}$ in the inner crust where $p_B\ll p$.
For $B_S=10^{12}$G, the eigenfunctions $iT_{l^\prime}$ and $ib^T_l$ are approximately
constant as functions of $r/R$, but for $B_S=10^{15}$G they are
influenced by the strong magnetic field in the outer crust where the magnetic pressure $p_B$
dominates the gas pressure $p$.

Reflecting the difference in the thickness of the solid crust between the two neutron star models,
the normalized frequency spectra of the toroidal torsional modes show substantial differences.
For example, the fundamental $_{l^\prime}t_0$ modes of the model NS13T8 have 
normalized frequencies $\bar\omega$ lower
than those of the model NS05T7, while the overtones of the former have higher $\bar\omega$
than the latter.
This property remains the same even in the presence of a strong magnetic field.
For the periods of the torsional modes of non-magnetic neutron stars, 
McDermott et al (1988) gave extrapolation formulae, in which the local periods are proportional to
the radius of the star for the fundamental modes and to the crust thickness for
the overtone modes for a given spherical harmonic degree $l^\prime$. 
The normalized frequencies $\bar\omega$ in the figure are consistent with what
the extrapolation formulae predict.

On a closer look at the behavior of the low frequency torsional modes 
of the model NS13T8 in Figure 1, one may find that the torsional modes suffer
mode crossings with magnetic modes, for which the oscillation frequency 
increases rapidly with increasing 
$B_S$ and the oscillation energy is dominated by the magnetic perturbations.
In fact, if we calculate the energies defined by
\be
E_K(r)={1\over 2}\int_0^r\omega^2\rho\left|\pmb{\xi}\right|^2dV, \quad\quad
E_B(r)={1\over 8\pi}\int_0^r\left|\pmb{B}^\prime\right|^2dV,
\ee
we find that $E_B(R)$ is much larger than $E_K(R)$ for the magnetic modes.
Note that $\nabla\cdot\pmb{\xi}=0$ for toroidal magnetic modes.
For more detail, see below \S 3.3.

\subsection{Spheroidal Modes}

Figure 3 plots the oscillation frequencies $\bar\omega$ of the low radial order
spheroidal shear modes $_{l=2}s_n$ and the core/crust interfacial modes $_li_2$ 
for the models NS05T7 (top panel) and NS13T8 (bottom panel).
For the model NS05T7, also plotted are magnetic modes, labeled $m_k$, 
that are found in the presence of 
a strong magnetic field.
Note that in this paper, we use the symbols $_li_1$ and $_li_2$ to denote the crust/ocean and
core/crust interfacial modes, respectively, where the amplitudes of
the former are strongly localized in a narrow region around the
crust/ocean interface and the amplitudes of the latter are 
largest at the core/crust interface.
The oscillation frequencies $\bar\omega$ of the spheroidal shear modes
increase with increasing $B_S$ and this frequency increase becomes more rapid for
higher radial order $n$ modes.
As in the case of toroidal torsional modes, for a given radial order $n$, the spheroidal shear
modes $_ls_n$ with different degrees $l$
are nearly degenerate in frequency, and mode couplings between the shear modes
become significant for strong $B_S$.
Figure 3 also shows that the frequency of the core/crust interfacial modes $_li_2$
is hardly modified by the magnetic field as strong as $B_S=10^{15}$G, except that
the interfacial modes of the model NS05T7 
expreience avoided crossings with magnetic modes, whose oscillation
frequencies rapidly increase with increasing $B_S$.

As in the case of the toroidal torsional modes,
the normalized frequency spectra of the spheroidal modes
of the two models appear differently, reflecting the
difference in the thickness of the crust.
The different appearance for the shear modes of the two models can be
understood by using an approximation formula for the period, which
is proportional to the crust thickness (e.g., McDermott et al 1988).

The eigenfunctions $H_l$ and $b^H_{l^\prime}$ of a $_2s_1$ mode of the model NS05T7
are plotted in Figure 4 for $B_S=10^{12}$G in panels (a) and (b) and 
for $B_S=10^{15}$G in panels (c) and (d), 
where the amplitude normalization is given by $S_{l_1}(R)=1$ at the surface.
Note that because of the jump conditions applied at the interfaces between the solid crust
and the fluid zones (see Appendix B), the horizontal component of the displacement vector is
continuous at the outer interface, but it is discontinuous at the inner interface, at which
almost free slippery jump condtions are employed.
For $B_S=10^{12}$G, the eigenfunctions $b^H_{l^\prime}$ exhibit zigzag behavior
in the outer crust reflecting that of $\Gamma_1$, which is caused by discontinuous change of
equilibrium composition with increasing $\rho$.
For $B_S=10^{12}$G, the expansion coefficients
$H_{l_1}$ and $b^H_{l^\prime_1}$ are dominant over others, 
but for $B_S=10^{15}$G the first few components of 
$H_{l}$ and $b^H_{l^\prime}$ are comparable to each other, and the eigenfunctions
in the outer crust are significantly modified by the strong field comaperd with
those for the case of $B_S=10^{12}$G.
As shown by Figure 4, the amplitude of $b^H_{l^\prime}$ for $B_S=10^{12}$G is much larger
than that for $B_S=10^{15}$G for the same amplitude normalization $S_{l_1}(R)=1$.
However, if we compare the quantities $E_K(r)$ and $E_B(r)$, we find
$E_K(R)$ is much larger than $E_B(R)$ for $B_S=10^{12}$G, but $E_K(R)$ and $E_B(R)$ are
comparable to each other for $B_S=10^{15}$G so that
the magnetic perturbations is important to determine the modal properties of the
shear modes in the presence of a strong magnetic field.

In Figure 5, the frequency $\bar\omega$ of the fundamental $_2f$ mode 
of the model NS05T7 is plotted versus $B_S$.
The $_2f$ mode heavily suffers mode crossings with high radial order shear modes $_ls_n$
as $B_S$ increases, and 
no pure $_2f$ mode could be identified in the presence of a strong magnetic field.
This is also the case for high frequency $p$ modes for a strongly magnetized star.
Note that, even at $B_S\sim10^{12}$G, 
the fundamental mode $_{2}f$ with $\bar\omega=1.8862$ is affected by the shear mode
$_{2}s_{n=10}$ that has the frequency $\bar\omega=1.8726$ at $B_S=0$.

The frequency $\bar\omega$ of the $g_1$ mode in the 
fluid ocean and that of the core/ocean interfacial mode $_{l=2}i_1$
are plotted versus $B_S$ for the model NS05T7 in Figure 6.
The frequencies begin to decrease rapidly to zero beyond
$B_S\sim 10^5$G, suggesting the strong magnetic field suppresses these modes.
This suppression may come from the dominance of the magnetic force over the buoyant force in the 
fluid ocean at large $B_S$.
We obtain almost the same suppression of the surface $g$ modes and the crust/ocean interfacial
modes for the model MS13T8, but they can survive much stronger magnetic field $B_S\sim10^{10}$G.
This may be because the fluid ocean of the hotter model NS13T8 is much thicker
and has a more developed radiative region, 
compared to the model NS05T7, giving  much stronger buoyant force for the modes.
It is interesting to note that
the result for the $g$ and $_li_1$ modes in the fluid ocean in this paper
is different from that obtained by Carroll et al (1986), who suggested that
the frequency of $g_1$ proportionally increases with increasing $B_S$ to be a
magnetically dominating mode, which they called $g/m$ modes.
The reason for the discrepancy is probably the difference in the geometry
of the models and the magnetic field configulation between the two calculations.

\subsection{Magnetic Modes}

For both toroidal and spheroidal modes, we find oscillation modes that can be
regarded as magnetic modes, for which the frequency $\bar\omega$ rapidly increases 
with increasing $B_S$, 
and the magnetic energy $E_B(R)$ is dominant over the kinetic one $E_K(R)$,
although its appearance looks different depending on the
model structure, particularly, on the crust thickness.
It is also important to note that there seem to exist critical strengths of the 
magnetic field beyong which magnetic modes are allowed to exist.
We find that the low frequency torsional modes of the model NS13T8 are often affected by
mode crossing with the magnetic modes, but that those of the model NS05T7
rarely suffer from the mode crossings within the
frequency and magnetic field strength ranges we investigated in this paper.
In the same parameter ranges as those for the toroidal torsional modes, 
we find magnetic modes
interacting with the core/crust interfacial modes $_li_2$ for the model NS05T7, but
no examples of the mode crossings of this kind are found for the model NS13T8.

As an example of the mode crossings between the torsional modes and 
magnetic modes found for the model NS13T8, we plot, in Figure 7,
$\bar\omega$ of the $_2t_0$ (left panel) and $_2t_1$ (right panel) modes versus $\log B_s$.
The panels are magnifications of the corresponding parts from Figure 1, and 
clearly indicate mode interactions with magnetic modes as $B_S$ increases.
It is interesting to note that between $B_S=10^{12}$G and $\sim 10^{13}$G, the $_2t_1$ mode
also interact with sequences of modes whose frequencies decrease as $B_S$ increases.
Figure 8 shows an example of the eigenfunctions of a toroidal magnetic mode
having $\bar\omega=0.01557$ at $B_S=10^{14.48}$G for NS13T8, where the amplitude
normalization is given by $iT_{l^\prime=2}=1$ at the surface.
For this mode, the expansion coefficients $iT_{l^\prime=2}$ and $ib^T_{l=1}$
are dominant, and the magnetic perturbation $ib^T_l$ shows almost discontinuous
change at the crust/ocean interface, and 
has almost negligible amplitudes in the fluid ocean.
Figure 9 is an example of the eigenfunctions of a spheroidal magnetic mode of 
$\bar\omega=0.07965$
at $Bs=10^{14.45}$G for the model NS05T7, where
the amplitude normalization is given by $S_{l=0}(R)=1$.
For this magnetic mode the first components 
$H_{\hat l=2}$ and $b^H_{l^\prime=1}$ of the series expansion
are not necessarily dominant over other components associated with higher $l$s.
Note that the relative thickness of the surface ocean of the model NS13T8 is
much larger than that of NS05T7.

\subsection{Effects of Slow Rotation}

For slow rotation, we may approximate
the oscillation frequency of axisymmetric ($m=0$) modes as
\be
\bar\omega\approx\bar\omega_0+C_2\bar\Omega^2,
\ee
where $\bar\omega_0$ denotes the oscillation frequency for the non-magnetic and non-rotating case.
Note that the linear term in $\Omega$ is proportional to $m$ and does not appear for
axisymmetric modes, and that no effects of rotational deformation 
of the equilibrium structure on the oscillation
are included in our calculation and 
hence the second order effects exclusively come from the Coriolis force, which is
dominant for low frequency modes with $\bar\omega\ltsim 1$ (e.g., Lee 1993).
We tabulate in Table 1 the coefficients $C_2$ for several toroidal and spheroidal modes including
torsional $_{l^\prime}t_n$ and shear $_ls_{n}$ modes for the models NS05T7 and NS13T8.
Note that
we have used the labels $_li_1$ and $_li_2$ to denote respectively the crust/ocean and core/crust 
interfacial modes for both the models.
For the model NS05T7, the frequency of the $_2i_1$ mode for $B_S=0$ is lower than 
that of the $_2i_2$, but for the model NS13T8 
the frequency of the $_2i_1$ mode is higher than that of the $_2i_2$ mode.
From the table, we find that the second order response of the fundamental toroidal modes to rotation 
is negative and small, but that of the $_{l^\prime}t_{n=1}$ is positive and rather large 
in the sense that the frequency correction $C_2\bar\Omega^2$ can be
comparable to $\bar\omega_0$ itself for $\bar\Omega\sim0.1$.
This is particularly the case for the toroidal modes of the model NS13T8.
In the same sense, the response of the low radial order spheroidal shear modes $_ls_n$ with low $l$
to rotation can be substantial as well, particularly for the model NS13T8.
The effects of slow rotation on axisymmetric $f$ and $p_1$ modes, however, are not significant, 
the result of which is consistent with the calculation 
by Saio (1981) for a polytropic star with the index
$N=3$ and $\Gamma_1=5/3$, who have however taken into account the effects of rotational deformation.

\begin{table*}
\centering
\begin{minipage}{140mm}
\caption{Coefficient $C_2$ for axisymmetric ($m=0$) modes for the models NS05T7 and NS13T8}
\begin{tabular}{@{}ccccccc@{}}
\hline
&& NS05T7 &&& NS13T8& \\
\noalign{\vskip 3pt}
Mode & $\bar\omega_0$ & {\rm Period} (ms) & $C_2$& $\bar\omega_0$ & {\rm Period} (ms) & $C_2$\\
\noalign{\vskip 3pt}
$_{2}t_0$ & 0.04019 & 18.6 &$-6.75\times10^{-2}$ & 0.01897 & 17.4 & $-1.26\times10^{-1}$\\
$_{2}t_1$ & 0.3286 & 2.29 & $~~2.76\times10^1$ & 0.4534 & 0.727 & $2.10\times10^{2}$\\
$_{3}t_0$ & 0.06352 & 11.8 & $-6.20\times10^{-2}$ & 0.03000 &11.0 & $-8.31\times10^{-2}$\\
$_{3}t_1$ & 0.3318 & 2.26 & $~~1.31\times10^1$ & 0.4539 & 0.726 & $1.05\times10^2$\\
$_{4}t_0$ & 0.08517 & 8.81 & $-9.83\times10^{-2}$ & 0.04024 & 8.19 &  $-6.39\times10^{-2}$\\
$_{4}t_1$ & 0.3361 & 2.23 & $~~6.74\times10^0$ & 0.4546 & 0.725 & $6.94\times10^1$\\
\noalign{\vskip 3pt}
$_{2}i_1$ & 0.008282 & 90.6 & $~~1.03\times10^2$ & 0.03326 & 9.91 & $2.56\times10^1$\\
$_{3}i_1$ & 0.01171 & 64.1 & $~~7.95\times10^1$ & 0.04703 & 7.01 & $1.92\times10^1$\\
$_{2}i_2$ & 0.1029 & 7.30 & $~~1.10\times10^1$ & 0.01862 & 17.7 & $1.51\times10^0$\\
$_{3}i_2$ & 0.1400 & 5.36 & $~~8.33\times10^0$ & 0.02889 & 11.4 & $2.52\times10^0$\\
\noalign{\vskip 3pt}
$_{2}s_1$ & 0.3093 & 2.43 & $-2.10\times10^1$ & 0.4512 & 0.731 & $-1.92\times10^2$ \\
$_{2}s_2$ & 0.5556 & 1.35 & $-5.37\times10^1$ & 0.7704 & 0.428 & $-1.84\times10^2$\\
$_{3}s_1$ & 0.2884 & 2.60 & $-2.10\times10^1$ & 0.4494 & 0.734 & $-1.03\times10^2$\\
$_{3}s_2$ & 0.5518 & 1.36 & $-9.51\times10^0$ & 0.7684 & 0.429 & $-1.02\times10^2$\\
\noalign{\vskip 3pt}
$_{2}f$ & 1.886 & 0.398 & $~~3.11\times10^{-1}$ & 1.434 & 0.230 & $1.39\times10{-1}$\\
$_{2}p_1$ & 4.038 & 0.186 & $~~2.09\times10^{-1}$ & 3.967 & 0.0831 & $1.42\times10^{-2}$\\
\hline
\end{tabular}
\end{minipage}
\end{table*}

\section{conclusions}

We have calculated axisymmetric $(m=0)$ oscillation modes of neutron star models that have
a solid crust and are threaded by a dipole
magnetic field.
We find that the frequencies of the fundamental toroidal torsional modes are 
not affected significantly by the magnetic field 
as strong as $B_S\sim10^{15}$G at the surface, and that
high radial order torsional and shear modes are susceptible to a magnetic field 
even if the strength at the surface is much less than $B_S\sim10^{15}$G.
Because both spheroidal shear and toroidal torsional modes are almost degenerate in
oscillation frequency for a given radial order $n$, the high radial order modes with
different $l$s for a given $n$ are easily coupled
in the presence of a strong magnetic field.
Since the $f$ modes (and $p$ modes) are embedded in the sea of high radial order shear modes 
with various $l$s that are sensitive to the
magnetic field, the $f$ modes suffer mode crossings with the shear modes as
$B_S$ varies, and their identity may become ambiguous in the presence of a strong magnetic field.

We find that the $g$ modes in the fluid ocean and the crust/ocean interfacial modes
are suppressed in a strong magnetic field, the result of which contradicts that
obtained by Carroll et al (1986), who showed that the ocean $g$ modes will survive to be
a magnetic mode, labeled $m/g_k$ in their paper, as the magnetic field becomes strong.
The reason for the contradiction may be attributable to the difference in the geometry of the 
oscillating stars and of the magnetic field, that is,
Carroll et al (1986) calculated oscillation modes of cylindrical stars threaded by 
a uniform magnetic field.
The strength of the magnetic field necessary to completely suppress the ocean $g$ modes
and the crust/ocean interfacial modes
depends on neutron star models, and if a neutron star has a hot buoyant radiative
region in the ocean, the modes can survive a magnetic field of $B_S\sim10^{10}$G or stronger,
although their frequency spectrum could be largely modified by the field.

As a model for burst oscillations observed in many low mass X-ray binary (LMXB) systems
(e.g., Strohmayer et al 1997, Strohmayer \& Bildsten 2004, van der Klis 2004), 
Heyl (2004) and Lee (2004) proposed that the oscillations are produced by
low frequency buoyant $r$ modes propagating in the surface fluid ocean of accreting neutron stars
in the systems (see also Lee \& Strohmayer 2005, Heyl 2005).
Since no effects of magnetic field are properly taken into account in their analyses, 
it is needed to reexamine the $r$ mode model for burst oscillations,
whether the buoyant $r$ modes can survive a
strong magnetic field of the neutron stars and how their frequency spectrum is modified by the field.

We have also examined the effects of slow rotation on the axisymmetric oscillation modes,
although no effects of rotational deformation are included.
Since the first order effects of rotation do not appear for axisymmetric modes with $m=0$,
it is the second order effects of rotation that appear first and are 
due to the Coriolis force when no rotational deformation is considered.
The second order effects can be
important for torsional modes and shear modes with radial order $n\ge1$
in the sense that the second order corrections to the frequency due to rotation
can be comparable to the frequency itself for $\bar\Omega\sim 0.1$.
But, for SGR 1806-20, for example, the rotation period
is estimated to be 7.56s (e.g., Israel et al 2005), implying that
the underlying object is a very slow rotator in the neutron star standard
so that $\bar\Omega\ll 1$ and $C_2\bar\Omega^2$ should be negligible
compared to $\bar\omega_0$.
The first order correction due to the Coriolis force 
for non-axisymmetric modes can be found in Lee \& Strohmayer (1996).

We find magnetic modes for a strong magnetic field.
Here, magnetic modes are regarded as a oscillation mode that 
exists only in the presence of a strong magnetic field, 
and whose frequency rapidly increases with increasing $B_S$.
We also note that the magnetic modes have their oscillation energy dominantly 
possessed by the magnetic perturbations.
The modal properties of the magnetic modes found by Carroll et al (1986), 
labeled $m/g_k$ and $a_k$, are not the same as the properties of 
the magnetic modes found in this paper.
Note that although the magnetic modes computed by Carroll et al (1986) reside in the fluid ocean,
the magnetic modes found in this paper
have amplitudes both in the fluid ocean and in the solid crust, and
a large fraction of the oscillation energy resides in the crust, as indicated by Figures 8 and 9.
We note that a simple magnetohydrodynamical system can support Alfv\'en waves, as well as
fast and slow magneto-acoustic waves
in fluids (e.g., Sturrock 1994), and we may expect almost the same modal structure 
even for a magnetic solid, in which torsional waves propagate (e.g., Carroll et al 1986).
At this moment, however, we have no clear classification scheme 
for the magnetic modes, which makes it difficult to obtain a good understanding of the modes.
Further studies are definitely necessary for magnatic modes of neutron stars with a solid crust,
since the magnetic modes could be important observationally for magnetor as an agent
triggering instability for flares.

Because of the assumption of non-magnetic core employed in this paper (\S 2.2), 
we completely ignore the possible existence of magnetosonic modes in the core and 
the possible coupling between core magnetic modes and crustal shear modes, for example.
This assumption would be a serious flaw when we are interested in core (or more global) 
magnetic modes themselves.
In fact, the crustal magnetic modes found in the present paper 
could be flawed in the sense that the assumption of non-magnetic core excludes 
from the beginning the possible existence of more global magnetic modes extending 
form the core to the crust.
For non-magnetic modes such as crustal shear modes, however,
so long as the field strength is less than $\sim 10^{15}$G at the surface so that
the magnetic pressure is much smaller than the gas pressure (and/or the shear modulus)
in the inner crust (see, e.g., Fig. 3 of Carroll et al (1986) or Fig. 1 of Piro (2005)),
reflection of crustal shear waves at the bottom of the crust is effective to establish 
crustal shear modes well trapped in the crust, and the effects of the coupling 
with core magnetic modes would be minor, for example, on the frequency spectrum 
of the crustal shear modes.
For the field strength much larger than $10^{15}$G, the confinement of 
crustal shear mode amplitudes into the crust would be imperfect and global treatment 
properly including the core will be necessary. As suggested by one of the referees, 
a differentially rotating magnetor progenitor would produce
a toroidal magnetic field, for which the complexities possibly
caused by mechanical coupling between the crust and the core could be avoided. 
This case may be important and worth careful examination.
Recent discussions on the crust-core coupling of magnetic modes, leading to 
more global magnetic modes of neutron stars, 
are found in Levin (2006), Glampedakis, Samuelsson, \& Andersson (2006), and 
Sotani, Kokkotas, \& Stergioulas (2006).

The assumption of non-magnetic core also leads to neglect of another possible role 
played by perturbed magnetic field in the core.
Several authors (e.g., Biront et al 1982, Roberts \& Soward 1983, Cambell \& Papaloizou 1986)
have discussed for magnetic normal stars that in a deep fluid region where
the magnetic pressure is much smaller than the gas pressure, magnetic perturbations 
become extremely short, decoupled from mechanical perturbations, which leads to a dissipation 
of the oscillation energy.
In this paper, however, because of the numerical difficulty in properly
treating the abrupt change of wave properties between the solid crust and the fluid core,
we have employed a simplifying assumption that the fluid core is non-magnetic.
Although the assumption of non-magnetic core would be reasonably justfied
since in the fluid core the magnetic pressure $p_B$ is much smaller than the gas pressure $p$,
appropriate estimations of the dissipative effects of extremely short magnetic perturbations 
in the fluid core
are necessary to see whether the dissipation is significant enough
to damp the excitation of the torsional oscillations.
We have almost the same difficulty in treating the outer boundary conditions, for which
we have assumed for simplicity no emission of electromagnetic waves from the surface even 
if we include
the displacement current term.
Although Carroll et al (1986) (see also McDermott et al 1988) suggested that
the damping effects due to emission of electromagnetic waves from the surface
are negligible except for magnetic modes, labeled $m/g_k$ and $a_k$, fully consistent
calculations including the damping effects are obviously necessary.

In the giant flare of SGR 1806-20 observed December 2004, there exist reports on
detection of QPOs at 18, 30, and 92.5 Hz (Israel et al 2005), 
at 18, 92.5, and 626.5 Hz (Watts \& Strohmayer 2006), and
at $\sim$90, $\sim$150, 625, and 1835 Hz, and at $\sim$720 and 2384 Hz but with less significances
(Strohmayer \& Watts 2006).
In the giant flare of SGR 1900+14 observed in August 1998, 
Strohmayer \& Watts (2005) have also found detection of QPOs at
53.5, 84, and 155.5 Hz, and at 28 Hz with lower significance.
Identifying the QPOs of low frequencies $\nu_\infty \ltsim 100$Hz 
with the fundamental toroidal torsionl modes with various $l$s
is rather secure (McDermott et al 1988, Duncan 1998, Piro 2005),
but identification of the middle to high frequency QPOs is not straightforward.
In fact, 
QPOs with frequencies $\sim100$ to $\sim1000$ Hz can be generated by high $l^\prime$ $_{l^\prime}t_0$
modes, or overtone modes $_{l^\prime}t_{n\ge1}$ with low $l^\prime$, or 
low $l$ spheroidal shear modes $_ls_{n\ge1}$,
or core/crust interfacial modes $_ls_2$, depending on neutron star models,
and there exist no unique solution to
identification if the frequency is the only available information.
For much higher frequency QPOs like that at 2384Hz, low $l$ fundamental modes $_lf$ 
must be added to the list of candidates for the QPOs, 
although much more energy will be required to excite the modes to
observable amplitudes than that for toroidal crustal modes.

In this paper, we have been only concerned with axisymmetric modes with $m=0$.
If we extend our calculation to non-axisymmetric modes with $m\not= 0$, 
mode coupling effetcs will be much more significant in a strong magnetic field
because both toroidal torsional and spheroidal shear modes are nearly degenerate in
frequency for a given radial order $n$, which makes numerical analysis
difficult and tedious.
However, together with employing neutron star models constructed with up to date
equations of state and shear modulus, the extension to
non-axisymmetric modes will be inevitable in order to make possible serious comparions between
theoretical predictions and observations.

\begin{appendix}

\section{Oscillation equations for a magnetized star for axisymmetric modes with $m=0$}

In this Appendix, we give the oscillation equations for axisymmetric ($m=0$) modes separately for 
even modes and for odd modes.
In the followings, for a given matrix $F=(F_{i,j})$, we employ
the symbols $\sdp{F}$, $\cvp{F}$, and $\kgp{F}$ to indicate
the matrices defined as
\be
\sdp{F}=(F_{i,j+1}), \quad \cvp{F}=(F_{i+1,j}), \quad \kgp{F}=(F_{i+1,j+1})
\ee
for $i,j=1,~2,~3,~\cdots$.

\subsection{Even Modes}

For the solid crust, we employ the dependent variables defined as
\be
\pmb{z}_1=\left(S_{l_j}(r)\right), \quad \pmb{z}_2=\left(H_{\hat l_j}(r)\right), 
\quad \pmb{z}_3=\left(iT_{l^\prime_j}(r)\right), \quad
\pmb{z}_4=\alpha_2\left({1\over r^2}{d\over dr}\left(r^3\pmb{z}_1\right)-\sdp{\Lambda_0}\pmb{z}_2\right)
+2\alpha_1{d\over dr}\left(r\pmb{z}_1\right)+{2p_B\over p}C_0\pmb{b}^H, 
\ee
\be
\pmb{z}_5=\alpha_1\left(r{d\pmb{z}_2\over dr}+\cvp{\unitm}\pmb{z}_1\right)-{4p_B\over p}\cvp{\pmb{M}_1}\pmb{b}^H, \quad
\pmb{z}_6=\alpha_1r{d\pmb{z}_3\over dr}-{4p_B\over p}\sdp{M_0}i\pmb{b}^T, \quad
\pmb{b}^H=\left(b^H_{l^\prime_j}(r)\right), 
\quad \pmb{b}^T=\left(b^T_{\hat l_j}(r)\right),
\ee
where
\be
l_j=2(j-1), \quad \hat l_j=2j, \quad l^\prime_j=1+2(j-1) \quad {\rm for} \quad j=1,~2,~3,~\cdots,
\ee
and the oscillation equations are given by
\be
r{d\pmb{z}_1\over dr}=-{3\Gamma_1\over\alpha_3}\pmb{z}_1+{\alpha_2\over\alpha_3}\sdp{\Lambda_0}\pmb{z}_2
+{1\over\alpha_3}\pmb{z}_4-{1\over\alpha_3}{2p_B\over p}C_0\pmb{b}^H,
\ee
\be
\sdp{M_0}r{d\pmb{z}_2\over dr}=2\left(1-{\alpha_1\over\alpha_3}\right)K\pmb{z}_1
+\left(\sdp{M_0}-{1\over 2}{\alpha_2\over\alpha_3}\sdp{C_1}\right)\pmb{z}_2
-{1\over2\alpha_3}K\pmb{z}_4+\left({1\over\alpha_3}{p_B\over p}KC_0-{1\over 2}\unitm\right)\pmb{b}^H,
\ee
\be
\cvp{\pmb{M}_1}r{d\pmb{z}_3\over dr}=\left(\cvp{\pmb{M}_1}-{1\over 2}\cvp{C_0}\right)\pmb{z}_3-{1\over 2}i\pmb{b}^T,
\ee
\begin{eqnarray}
r{d\pmb{z}_4\over dr}=\left(\left(U-4-c_1\bar\omega^2\right)V\unitm-{2p_B\over p}C_0\left(2Q_1+C_1\right)\right)\pmb{z}_1
+\left(\left(V-2\alpha_1\right)\sdp{\Lambda_0}-{4p_B\over p}C_0\sdp{Q_1\Lambda_0+C_1}\right)\pmb{z}_2
+c_1\bar\omega^2\nu VC_0\pmb{z}_3  \nonumber \\
+V\pmb{z}_4+\sdp{\Lambda_0}\pmb{z}_5-4\alpha_1r{d\pmb{z}_1\over dr}+{4p_B\over p}
\left(Q_0\Lambda_1-C_0\right)\pmb{b}^H 
+c_1\bar\omega^2V\left({v_A\over c}\right)^2\left(\left(Q_0Q_1-\pmb{1}\right)\pmb{z}_1
+2\left[Q_0C_1\right]\pmb{z}_2\right),
\end{eqnarray}
\begin{eqnarray}
r{d\pmb{z}_5\over dr}=\left(V\cvp{\unitm}+{4p_B\over p}\cvp{\pmb{M}_1}\left(2Q_1+C_1\right)\right)\pmb{z}_1+
\left(-c_1\bar\omega^2V\unitm-2\alpha_1\unitm+2\alpha_1\kgp{\Lambda_0}+{8p_B\over p}\cvp{\pmb{M}_1}\sdp{Q_1\Lambda_0+C_1}\right)\pmb{z}_2
\nonumber \\
+c_1\bar\omega^2\nu V\cvp{\pmb{M}_1}\pmb{z}_3-\cvp{\unitm}\pmb{z}_4+\left(V-3\right)\pmb{z}_5
+\cvp{\unitm}2\alpha_1r{d\pmb{z}_1\over dr}+{4p_B\over p}\left(\cvp{\pmb{M}_1}+{1\over2}\cvp{C_0}\right)\pmb{b}^H \nonumber\\
+c_1\bar\omega^2V\left({v_A\over c}\right)^2\kgp{\Lambda_0}^{-1}\left(-2\left(3Q_0Q_1-1+Q_0C_1\right)\pmb{z}_1
-\left[4Q_0Q_1\Lambda_0+8Q_0C_1\right]\pmb{z}_2\right),
\end{eqnarray}
\begin{eqnarray}
r{d\pmb{z}_6\over dr}=-c_1\bar\omega^2\nu VK\pmb{z}_1+c_1\bar\omega^2\nu V\sdp{M_0}\pmb{z}_2
+\left(-c_1\bar\omega^2V\unitm-2\alpha_1(\unitm-{1\over 2}\Lambda_1)\right)\pmb{z}_3+\left(V-3\right)\pmb{z}_6
+{4p_B\over p}\left(\sdp{M_0}+{1\over 2}\sdp{C_1}\right)i\pmb{b}^T\nonumber\\
-c_1\bar\omega^2V\left({v_A\over c}\right)^2\Lambda_1^{-1}
\left(6Q_1C_0+4Q_1Q_0\Lambda_1-C_1C_0\right)\pmb{z}_3,
\end{eqnarray}
where $\unitm$ denotes the unit matrix, and
\be
\bar\omega={\omega/\sqrt{GM/R^3}}, \quad \nu={2\Omega/ \omega},
\ee
and
\be
c_1={(r/R)^3\over M_r/M},
\quad V=-{d\ln p\over d\ln r}, \quad U={d\ln M_r\over d\ln r},  
\quad p_B={B_0^2(r)\over 8\pi}, \quad v_A=\sqrt{B_0^2(r)\over 4\pi\rho}
\ee
and
\be
\alpha_1={\mu\over p}, \quad \alpha_2=\Gamma_1-{2\over 3}\alpha_1, 
\quad \alpha_3=\Gamma_1+{4\over 3}\alpha_1,
\ee
and $M$ and $R$ are the mass and radius of the star, and $G$ is the gravitational constant.
Note that the terms proportional to $(v_A/c)^2$ come from the displacement current.

For fluid regions the dependent variables we use are defined as
\be
\pmb{y}_1=\pmb{z}_1,  \quad \pmb{y}_2=\left({p^\prime_{l_j}(r)\over pV}\right), 
\quad \pmb{y}_3=\pmb{z}_2, \quad
\pmb{y}_4=\pmb{z}_3, \quad \pmb{y}_5={4p_B\over pV}\pmb{b}^H, 
\quad \pmb{y}_6={4p_B\over pV}i\pmb{b}^T,
\ee
and the oscillation equations are given by
\be
r{d\pmb{y}_1\over dr}=\left({V\over\Gamma_1}-3\right)\pmb{y}_1
-{V\over\Gamma_1}\pmb{y}_2+\sdp{\Lambda_0}\pmb{y}_3,
\ee
\begin{eqnarray}
r{d\pmb{y}_2\over dr}=\left(\left(c_1\bar\omega^2+rA\right)\unitm+{2p_B\over pV}C_0\left(2Q_1+C_1\right)\right)\pmb{y}_1
+\left(1-rA-U\right)\pmb{y}_2+{4p_B\over pV}C_0\sdp{Q_1\Lambda_0+C_1}\pmb{y}_3-c_1\bar\omega^2\nu C_0\pmb{y}_4
\nonumber \\
+{f\over 2}C_0\pmb{y}_5+{C_0\over 2}r{d\pmb{y}_5\over dr}
-c_1\bar\omega^2\left({v_A\over c}\right)^2\left(\left(Q_0Q_1-\pmb{1}\right)\pmb{y}_1
+2\left[Q_0C_1\right]\pmb{y}_3\right),
\end{eqnarray}
\be
\sdp{M_0}r{d\pmb{y}_3\over dr}=-{1\over 2}\left({V\over\Gamma_1}-4\right)K\pmb{y}_1+{1\over 2}{V\over\Gamma_1}K\pmb{y}_2
+\left(\sdp{M_0}-{1\over 2}\sdp{C_1}\right)\pmb{y}_3-{1\over 2}\left({4p_B\over pV}\right)^{-1}\pmb{y}_5,
\ee
\be
\cvp{\pmb{M}_1}r{d\pmb{y}_4\over dr}=\left(\cvp{\pmb{M}_1}-{1\over 2}\cvp{C_0}\right)\pmb{y}_4
-{1\over 2}\left({4p_B\over pV}\right)^{-1}\pmb{y}_6,
\ee
\begin{eqnarray}
\cvp{\pmb{M}_1}r{d\pmb{y}_5\over dr}=-{4p_B\over pV}\cvp{\pmb{M}_1}\left(2Q_1+C_1\right)\pmb{y}_1
-\cvp{\unitm}\pmb{y}_2+\left(c_1\bar\omega^2\unitm-{8p_B\over pV}\cvp{\pmb{M}_1}\sdp{Q_1\Lambda_0+C_1}\right)\pmb{y}_3
-c_1\bar\omega^2\nu \cvp{\pmb{M}_1}\pmb{y}_4
\nonumber \\
-f\cvp{\pmb{M}_1}\pmb{y}_5
-c_1\bar\omega^2\left({v_A\over c}\right)^2\kgp{\Lambda_0}^{-1}\left(-2\left(3Q_0Q_1-1+Q_0C_1\right)\pmb{y}_1
-\left[4Q_0Q_1\Lambda_0+8Q_0C_1\right]\pmb{y}_3\right),
\end{eqnarray}
\begin{eqnarray}
\sdp{M_0}r{d\pmb{y}_6\over dr}=c_1\bar\omega^2\nu K\pmb{y}_1-c_1\bar\omega^2\nu\sdp{M_0}\pmb{y}_3
+c_1\bar\omega^2\pmb{y}_4-\left(f\sdp{M_0}+{1\over 2}\sdp{C_1}\right)\pmb{y}_6\nonumber\\
+c_1\bar\omega^2\left({v_A\over c}\right)^2\Lambda_1^{-1}\left(6Q_1C_0+4Q_1Q_0\Lambda_1-C_1C_0\right)\pmb{y}_4,
\end{eqnarray}
where
\be
f=rA-{V\over\Gamma_1}+U+3.
\ee

The non-zero elements of the matrices $C_0$, $C_1$, $K$, $M_0$, $\pmb{M}_1$, $Q_0$, $Q_1$, $\Lambda_0$, 
and $\Lambda_1$ that appear in the above equations are given by
\be
(Q_0)_{i,i}=J^m_{l+1}, \quad (Q_0)_{i+1,i}=J^m_{l+2}, \quad 
(Q_1)_{i,i}=J^m_{l+1}, \quad (Q_1)_{i,i+1}=J^m_{l+2},
\ee
\be
(C_0)_{i,i}=-(l+2)J^m_{l+1}, \quad (C_0)_{i+1,i}=(l+1)J^m_{l+2}, \quad 
(C_1)_{i,i}=lJ^m_{l+1}, \quad (C_1)_{i,i+1}=-(l+3)J^m_{l+2},
\ee
\be
(K)_{i,i}={J^m_{l+1}\over l+1}, \quad (K)_{i,i+1}=-{J^m_{l+2}\over l+2},
\ee
\be
(\Lambda_0)_{i,i}=l(l+1), \quad (\Lambda_1)_{i,i}=(l+1)(l+2),
\ee
\be
(M_0)_{i,i}={l\over l+1}J^m_{l+1}, \quad (M_0)_{i,i+1}={l+3\over l+2}J^m_{l+2}, \quad
(\pmb{M}_1)_{i,i}={l+2\over l+1}J^m_{l+1}, \quad (\pmb{M}_1)_{i+1,i}={l+1\over l+2}J^m_{l+2},
\ee
where $l=|m|+2i-2$ for $i=1,~2,~3,~\cdots$, and 
\be
J^m_{l}\equiv\left[{(l+m)(l-m)\over (2l-1)(2l+1)}\right]^{1/2}.
\ee

\subsection{Odd Modes}

For the crust, we use the dependent variables defined as
\be
\pmb{z}_1=\left(S_{l_j}(r)\right), \quad \pmb{z}_2=\left(H_{l_j}(r)\right), 
\quad \pmb{z}_3=\left(iT_{\hat l^\prime_j}(r)\right), \quad
\pmb{z}_4=\alpha_2\left({1\over r^2}{d\over dr}\left(r^3\pmb{z}_1\right)-\Lambda_0\pmb{z}_2\right)
+2\alpha_1{d\over dr}\left(r\pmb{z}_1\right)+{2p_B\over p}\sdp{C_0}\pmb{b}^H, 
\ee
\be
\pmb{z}_5=\alpha_1\left(r{d\pmb{z}_2\over dr}+\pmb{z}_1\right)-{4p_B\over p}\sdp{\pmb{M}_1}\pmb{b}^H, \quad
\pmb{z}_6=\alpha_1r{d\pmb{z}_3\over dr}-{4p_B\over p}\cvp{M_0}i\pmb{b}^T, \quad
\pmb{b}^H=\left(b^H_{\hat l^\prime_j}(r)\right), 
\quad \pmb{b}^T=\left(b^T_{l_j}(r)\right),
\ee
where
\be
l_j=2(j-1)+1, \quad \hat l^\prime_j=2j, \quad l^\prime_j=2(j-1) \quad {\rm for} \quad j=1,~2,~3,~\cdots.
\ee
The oscillation equations are then
\be
r{d\pmb{z}_1\over dr}=-{3\Gamma_1\over\alpha_3}\pmb{z}_1+{\alpha_2\over\alpha_3}\Lambda_0\pmb{z}_2
+{1\over\alpha_3}\pmb{z}_4-{1\over\alpha_3}{2p_B\over p}\sdp{C_0}\pmb{b}^H,
\ee
\be
\cvp{M_0}r{d\pmb{z}_2\over dr}=2\left(1-{\alpha_1\over\alpha_3}\right)\cvp{K}\pmb{z}_1
+\left(\cvp{M_0}-{1\over 2}{\alpha_2\over\alpha_3}\cvp{C_1}\right)\pmb{z}_2
-{1\over2\alpha_3}\cvp{K}\pmb{z}_4+\left({1\over\alpha_3}{p_B\over p}\cvp{K}\sdp{C_0}
-{1\over 2}\unitm\right)\pmb{b}^H,
\ee
\be
\sdp{\pmb{M}_1}r{d\pmb{z}_3\over dr}=\left(\sdp{\pmb{M}_1}-{1\over 2}\sdp{C_0}\right)\pmb{z}_3
-{1\over 2}i\pmb{b}^T,
\ee
\begin{eqnarray}
r{d\pmb{z}_4\over dr}=\left(\left(U-4-c_1\bar\omega^2\right)V\unitm-{2p_B\over p}C_0\left(2Q_1+C_1\right)\right)\pmb{z}_1
+\left(\left(V-2\alpha_1\right)\Lambda_0-{4p_B\over p}C_0\left(Q_1\Lambda_0+C_1\right)\right)\pmb{z}_2
+c_1\bar\omega^2\nu V\sdp{C_0}\pmb{z}_3\nonumber \\
+V\pmb{z}_4+\Lambda_0\pmb{z}_5-4\alpha_1r{d\pmb{z}_1\over dr}+{4p_B\over p}
\sdp{Q_0\Lambda_1-C_0}\pmb{b}^H
+c_1\bar\omega^2V\left({v_A\over c}\right)^2\left(\left(Q_0Q_1-\pmb{1}\right)\pmb{z}_1
+2Q_0C_1\pmb{z}_2\right),
\end{eqnarray}
\begin{eqnarray}
r{d\pmb{z}_5\over dr}=\left(V\unitm+{4p_B\over p}\pmb{M}_1\left(2Q_1+C_1\right)\right)\pmb{z}_1+
\left(-c_1\bar\omega^2V\unitm-2\alpha_1\left(\unitm-\Lambda_0\right)+{8p_B\over p}\pmb{M}_1\left(Q_1\Lambda_0+C_1\right)\right)\pmb{z}_2
\nonumber \\
+c_1\bar\omega^2\nu V\sdp{\pmb{M}_1}\pmb{z}_3-\pmb{z}_4+\left(V-3\right)\pmb{z}_5
+2\alpha_1r{d\pmb{z}_1\over dr}+{4p_B\over p}\left(\sdp{\pmb{M}_1}+{1\over2}\sdp{C_0}\right)\pmb{b}^H\nonumber\\
+c_1\bar\omega^2V\left({v_A\over c}\right)^2{\Lambda_0}^{-1}\left(-2\left(3Q_0Q_1-1+Q_0C_1\right)\pmb{z}_1
-\left(4Q_0Q_1\Lambda_0+8Q_0C_1\right)\pmb{z}_2\right),
\end{eqnarray}
\begin{eqnarray}
r{d\pmb{z}_6\over dr}=-c_1\bar\omega^2\nu V\cvp{K}\pmb{z}_1+c_1\bar\omega^2\nu V\cvp{M_0}\pmb{z}_2
+\left(-c_1\bar\omega^2V\unitm-2\alpha_1\unitm+\alpha_1\kgp{\Lambda_1}\right)\pmb{z}_3+\left(V-3\right)\pmb{z}_6
\nonumber\\
+{4p_B\over p}\left(\cvp{M_0}+{1\over 2}\cvp{C_1}\right)i\pmb{b}^T
-c_1\bar\omega^2V\left({v_A\over c}\right)^2\kgp{\Lambda_1}^{-1}
\left[6Q_1C_0+4Q_1Q_0\Lambda_1-C_1C_0\right]\pmb{z}_3.
\end{eqnarray}

For fluid regions the dependent variables we use are
\be
\pmb{y}_1=\pmb{z}_1,  \quad \pmb{y}_2=\left({p^\prime_{l_j}(r)\over pV}\right), 
\quad \pmb{y}_3=\pmb{z}_2, \quad
\pmb{y}_4=\pmb{z}_3, \quad \pmb{y}_5={4p_B\over pV}\pmb{b}^H, 
\quad \pmb{y}_6={4p_B\over pV}i\pmb{b}^T,
\ee
and the oscillation equations are then given by
\be
r{d\pmb{y}_1\over dr}=\left({V\over\Gamma_1}-3\right)\pmb{y}_1
-{V\over\Gamma_1}\pmb{y}_2+\Lambda_0\pmb{y}_3,
\ee
\begin{eqnarray}
r{d\pmb{y}_2\over dr}=\left(\left(c_1\bar\omega^2+rA\right)\unitm+{2p_B\over pV}C_0\left(2Q_1+C_1\right)\right)\pmb{y}_1
+\left(1-rA-U\right)\pmb{y}_2+{4p_B\over pV}C_0\left(Q_1\Lambda_0+C_1\right)\pmb{y}_3-c_1\bar\omega^2\nu \sdp{C_0}\pmb{y}_4
\nonumber \\
+{f\over 2}\sdp{C_0}\pmb{y}_5+{\sdp{C_0}\over 2}r{d\pmb{y}_5\over dr}
-c_1\bar\omega^2\left({v_A\over c}\right)^2\left(\left(Q_0Q_1-\pmb{1}\right)\pmb{y}_1
+2Q_0C_1\pmb{y}_3\right),
\end{eqnarray}
\be
\cvp{M_0}r{d\pmb{y}_3\over dr}=-{1\over 2}\left({V\over\Gamma_1}-4\right)\cvp{K}\pmb{y}_1+{1\over 2}{V\over\Gamma_1}\cvp{K}\pmb{y}_2
+\left(\cvp{M_0}-{1\over 2}\cvp{C_1}\right)\pmb{y}_3-{1\over 2}\left({4p_B\over pV}\right)^{-1}\pmb{y}_5,
\ee
\be
\sdp{\pmb{M}_1}r{d\pmb{y}_4\over dr}=\left(\sdp{\pmb{M}_1}-{1\over 2}\sdp{C_0}\right)\pmb{y}_4
-{1\over 2}\left({4p_B\over pV}\right)^{-1}\pmb{y}_6,
\ee
\begin{eqnarray}
\sdp{\pmb{M}_1}r{d\pmb{y}_5\over dr}=-{4p_B\over pV}\pmb{M}_1\left(2Q_1+C_1\right)\pmb{y}_1
-\pmb{y}_2+\left(c_1\bar\omega^2\unitm-{8p_B\over pV}\pmb{M}_1(Q_1\Lambda_0+C_1)\right)\pmb{y}_3
-c_1\bar\omega^2\nu \sdp{\pmb{M}_1}\pmb{y}_4-f\sdp{\pmb{M}_1}\pmb{y}_5\nonumber\\
-c_1\bar\omega^2\left({v_A\over c}\right)^2{\Lambda_0}^{-1}\left(-2\left(3Q_0Q_1-1+Q_0C_1\right)\pmb{y}_1
-\left(4Q_0Q_1\Lambda_0+8Q_0C_1\right)\pmb{y}_3\right),
\end{eqnarray}
\begin{eqnarray}
\cvp{M_0}r{d\pmb{y}_6\over dr}=c_1\bar\omega^2\nu \cvp{K}\pmb{y}_1-c_1\bar\omega^2\nu\cvp{M_0}\pmb{y}_3
+c_1\bar\omega^2\pmb{y}_4-\left(f\cvp{M_0}+{1\over 2}\cvp{C_1}\right)\pmb{y}_6\nonumber\\
+c_1\bar\omega^2\left({v_A\over c}\right)^2\kgp{\Lambda_1}^{-1}
\left[6Q_1C_0+4Q_1Q_0\Lambda_1-C_1C_0\right]\pmb{y}_4.
\end{eqnarray}

The non-zero elements of the matrices $C_0$, $C_1$, $K$, $M_0$, $\pmb{M}_1$, $Q_0$, $Q_1$, $\Lambda_0$, 
and $\Lambda_1$ are given by
\be
(Q_0)_{i,i}=J^m_{l}, \quad (Q_0)_{i,i+1}=J^m_{l+1}, \quad
(Q_1)_{i,i}=J^m_{l}, \quad (Q_1)_{i+1,i}=J^m_{l+1},  
\ee
\be
(C_0)_{i,i}=(l-1)J^m_{l}, \quad (C_0)_{i,i+1}=-(l+2)J^m_{l+1}, \quad 
(C_1)_{i,i}=-(l+1)J^m_{l}, \quad (C_1)_{i,i+1}=(l)J^m_{l+1},
\ee
\be
(K)_{i,i}=-{J^m_{l}\over l}, \quad (K)_{i,i+1}={J^m_{l+1}\over l+1},
\ee
\be
(\Lambda_0)_{i,i}=l(l+1), \quad (\Lambda_1)_{i,i}=(l-1)l,
\ee
\be
(M_0)_{i,i}={l+1\over l}J^m_{l}, \quad (M_0)_{i+1,i}={l\over l+1}J^m_{l+1}, \quad
(\pmb{M}_1)_{i,i}={l-1\over l}J^m_{l}, \quad (\pmb{M}_1)_{i+1,i}={l+2\over l+1}J^m_{l+1},
\ee
where $l=|m|+2i-1$ for $i=1,~2,~3,~\cdots$.

Note that for axisymmetric modes with $m=0$, the components $H_{l_1}$ and $b^T_{l_1}$ for even modes
and $iT_{l^\prime_1}$ and $b^H_{l^\prime_1}$ for odd modes are missing.
As is evident from the equations given above, if $\Omega=0$, the oscillation equations for axisymmetric
modes are separated into those for spheroidal and toroidal modes, respectively.
In this case, we can obtain spheroidal modes and toroidal modes separately, and mode couplings between
these two are brought about as a result of the effects of rotation.

\section{Boundary Conditons and Jump Conditions}

The surface boundary conditions we use are the same as those given in Lee (2005), and they are
at the stellar surface
\be
\delta p/p=0, \quad \pmb{b}^S+L^+\pmb{b}^H=0,
\ee
and
\be
i\pmb{b}^T=0, 
\ee
where $(L^+)_{ij}=\delta_{ij}(l^\prime_j+1)$,
$\delta p$ is the Lagrangian perturbation of the pressure, and
\be
\pmb{b}^S\equiv (b^S_{l^\prime_j}(r))=-\Lambda_1K\pmb{z}_1-2\Lambda_1M_0\pmb{z}_2-2m\pmb{z}_3.
\ee
Assuming the central region is a non-magnetic fluid core, the central boundary conditions we impose
are the regularity conditions for the variables $r\pmb{y}_1$ and $r\pmb{y}_2$
at the center (see, e.g., Lee \& Saio 1986).

For the jump conditions at the interfaces between the solid crust and fluid regions,
we assume that the dipole magnetic fiels is continuous at the interfaces, and that there appears
no singularity in $\pmb{B}^\prime$ at the interfaces.
Then, with the continuity of the radial component of the displacement vector given by
\be
\pmb{z}_1=\pmb{y}_1, 
\ee 
the conditions 
\be
\left[\left(\pmb{\xi}\times\pmb{B}\right)\cdot\pmb{e}_\theta\right]_-^+=0, \quad
\left[\left(\pmb{\xi}\times\pmb{B}\right)\cdot\pmb{e}_\phi\right]_-^+=0,
\ee
lead to
\be
\pmb{z}_2=\pmb{y}_3,
\ee
and
\be
\pmb{z}_3=\pmb{y}_4,
\ee
where $\left[F\right]_-^+\equiv\lim_{\epsilon\rightarrow 0}\left[F(r+\epsilon)-F(r-\epsilon)\right]$.
Note that the conditions $(B5)$ are obtained by
integrating the right hand side of equation (5) along a closed, infinitisimally small, curve that crosses the interface.
The conditions $(B4)$, $(B6)$, and $(B7)$ guarantee the continuity of the radial component of $\pmb{B}^\prime$ at the
interfaces because of equation $(B3)$.
Finally, we require the continuity of the three components of traction given by
\be
\left[\delta\tau^B_{rr}+\delta\tau^S_{rr}\right]^+_-=0, 
\ee
\be
\left[\delta\tau^B_{r\theta}+\delta\tau^S_{r\theta}\right]^+_-=0, 
\ee
and
\be
\left[\delta\tau^B_{r\phi}+\delta\tau^S_{r\phi}\right]^+_-=0,
\ee
where $\delta\tau^S_{ij}$ and $\delta\tau^B_{ij}$ are the $ij$ components of the perturbed traction associated with 
bulk modulus and magnetic field, respectively, and they are given by
\be
\delta\tau^S_{rr}=\sum_{l}\left\{\alpha_2\left[{1\over r^2}{d\over dr}(r^3S_l)-l(l+1)H_l\right]
+2\alpha_1{d\over dr}(rS_l)\right\}Y_l^m,
\ee
\be
\delta\tau^S_{r\theta}=\alpha_1\sum_{l,l^\prime}\left[\left(r{dH_l\over dr}+S_l\right){\partial Y^m_{l}\over
\partial\theta}
+r{dT_{l^\prime}\over dr}{1\over\sin\theta}{\partial Y^m_{l^\prime}\over\partial\phi}\right],
\ee
\be
\delta\tau^S_{r\phi}=\alpha_1\sum_{l,l^\prime}\left[\left(r{dH_l\over dr}+S_l\right)
{1\over\sin\theta}{\partial Y^m_l\over\partial\phi}+r{dT_{l^\prime}\over dr}{\partial Y^m_{l^\prime}\over
\partial\theta}\right],
\ee
\be
\delta\tau^B_{rr}={1\over 4\pi}\left(B_r\delta B_r-B_\theta\delta B_\theta\right),
\ee
\be
\delta\tau^B_{r\theta}={1\over 4\pi}\left(B_\theta\delta B_r+B_r\delta B_\theta\right),
\ee
and 
\be
\delta\tau^B_{r\phi}={1\over 4\pi}B_r\delta B_\phi.
\ee
Note that 
\be
\left[B_i\delta B_j\right]^+_-
=\left[B_i\left(B_j^\prime+\pmb{\xi}\cdot\nabla B_j\right)\right]^+_-=\left[B_iB_j^\prime\right]^+_-,
\ee
since $\left[\pmb{\xi}\right]^+_-=0$ and $\left[\pmb{B}\right]^+_-=0$ at the interface.

In this paper, all the jump conditions discussed above are applied at the interface between the
solid crust and the fluid ocean.
However, since we have assumed the fluid core extending below the solid crust is non-magnetic,
we cannot use all the jump conditions
at the interface between the solid crust and the fluid core. 
Here, we apply the jump conditions (B4), (B8), (B9), and (B10).
In the limit of non-magnetized stars, these jump conditions reduce to free slippery conditions
employed by McDermott et al (1988), and Lee \& Strohmayer (1996).
Because of the jump conditions we use, the horizontal and toroidal components
of the displacement vector and the perturbed magnetic field are not necessarily
continuous at the interface.

\end{appendix}



\newpage

\begin{figure}
\centering
\psfig{file=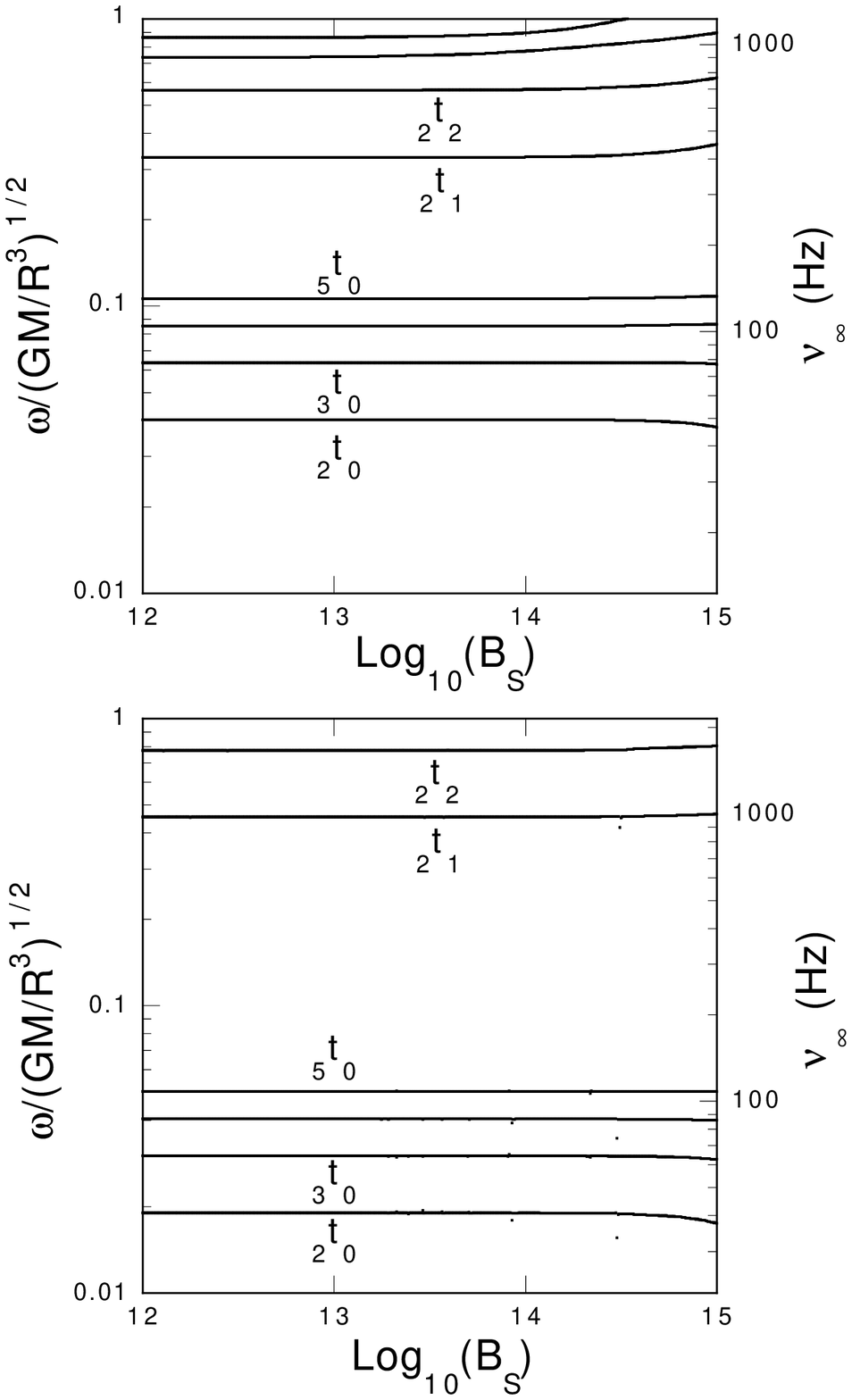,width=0.8\textwidth}
\caption{ $\bar\omega$ versus $\log(B_S)$ for axisymmetric ($m=0$) toroidal torsional
modes $_{l^\prime}t_n$ of the neutron star models NS05T7 (top panel)
and NS13T8 (bottom panel), where the fundamental torsional modes $_{l^\prime}t_{n=0}$
for $l^\prime=2$ to 5,
and the overtone modes $_{l^\prime=2}t_{n}$ for $n=1$ to 4 for NT05T7 and
for $n=1$ and 2 for NS13T8 are shown.
The frequency $\bar\omega$ shown is the local one, and 
$\nu_\infty=\omega\left(1-2GM/Rc^2\right)^{1/2}/2\pi$ is also dsiplayed on the right axis
of each of the panels.}
\end{figure}

\begin{figure}
\centering
\psfig{file=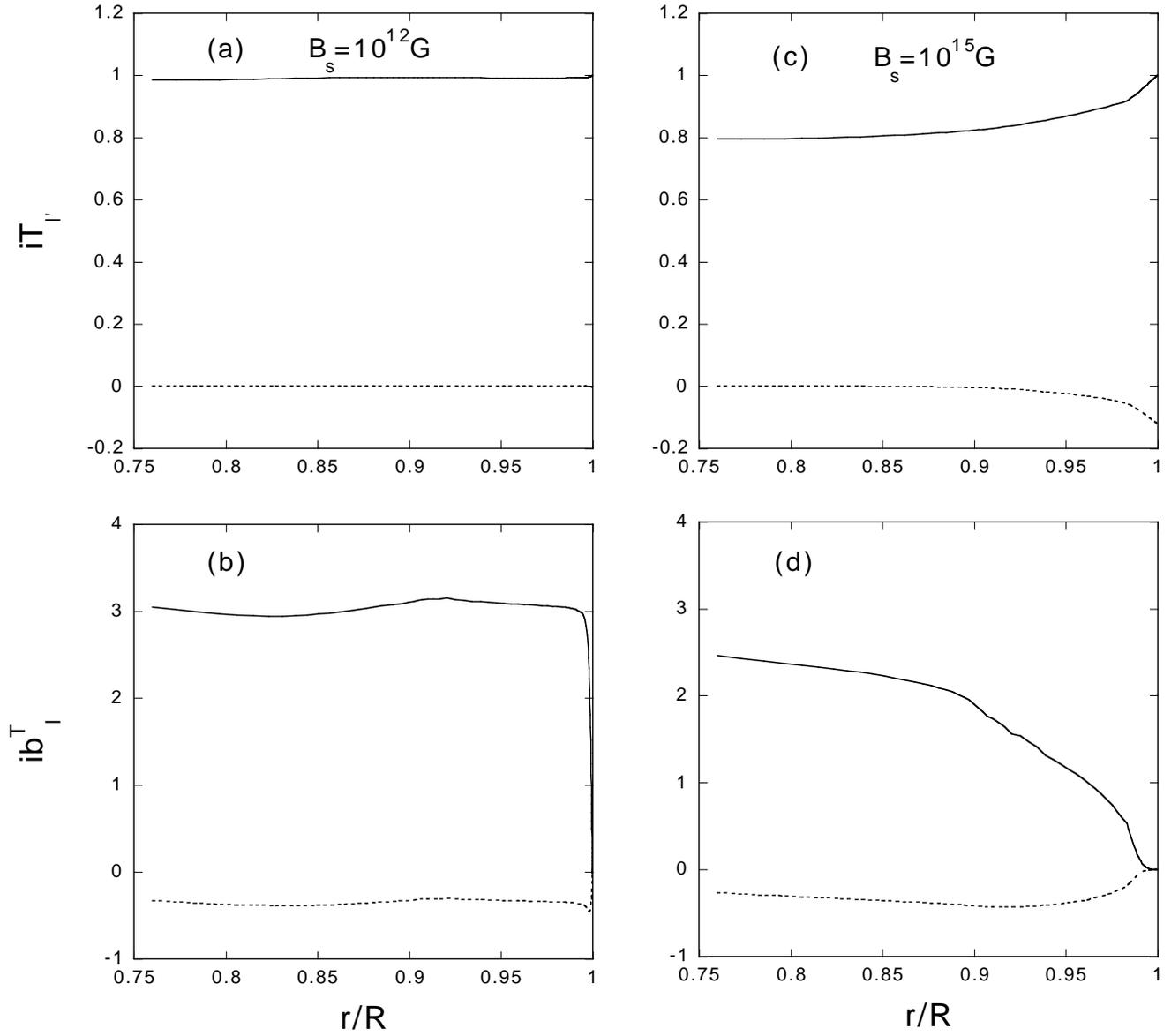,width=1.\textwidth}
\caption{Eigenfunctions $iT_{l^\prime}$ and $ib^T_{l}$ of the axisymmetric ($m=0$) fundamental
toroidal torsional mode $_2t_0$ for $B_S=10^{12}$G in panels (a) and (b)
and for $B_S=10^{15}$G in panels (c) and (d), where
the solid lines stand for $iT_{l^\prime=2}$ and $ib^T_{l=1}$, and
the dashed lines for $iT_{l^\prime=4}$ and $ib^T_{l=3}$, respectively.
The amplitude normalization is given by $iT_{l^\prime=2}=1$ at the surface.
}
\end{figure}

\begin{figure}
\centering
\psfig{file=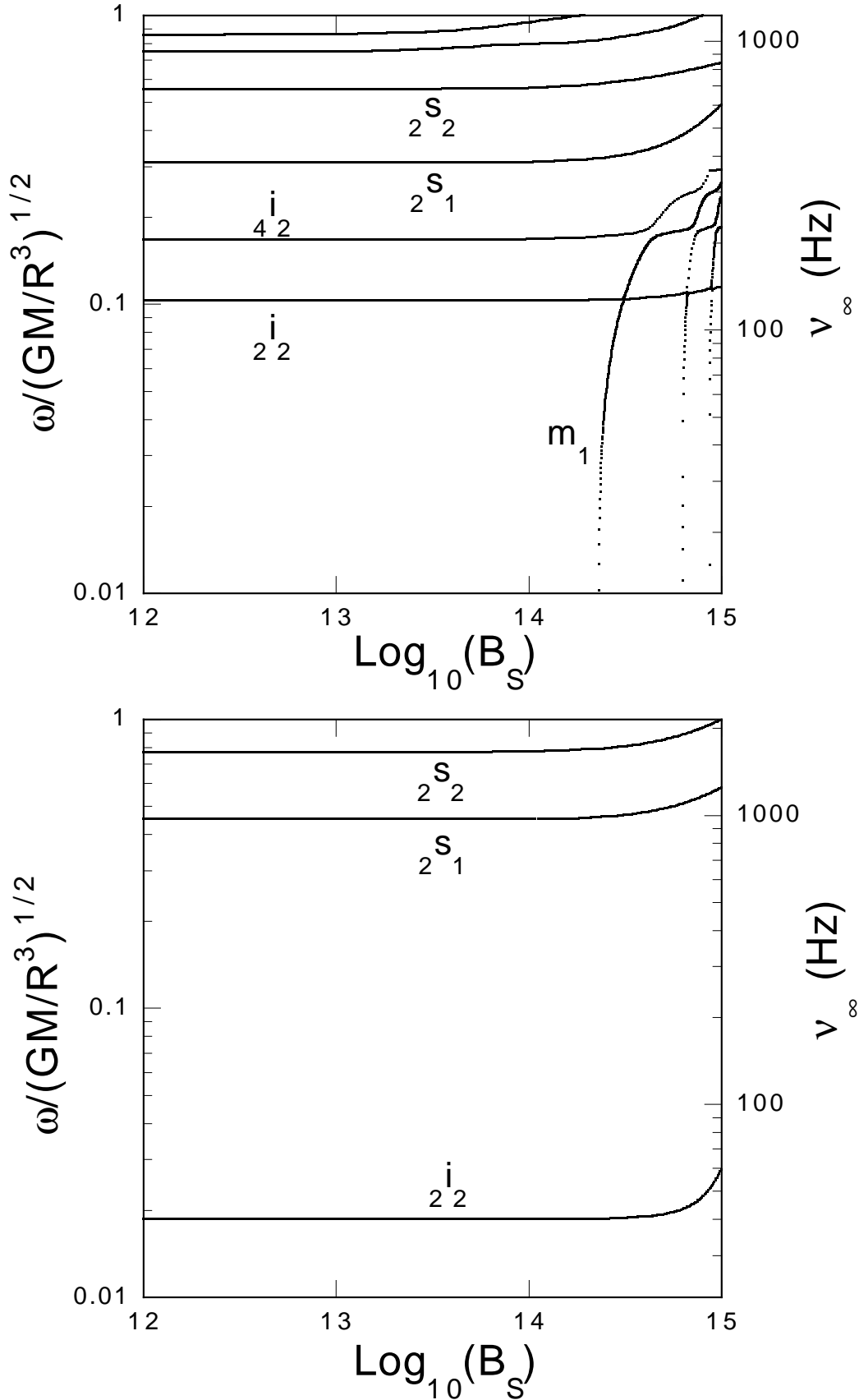,width=0.8\textwidth}
\caption{$\bar\omega$ versus $\log(B_S)$ for axisymmetric ($m=0$) spheroidal shear modes 
$_{l}s_n$ and core/crust interfacial modes $_li_2$
of the neutron star models NS05T7 (top panel) and NS13T8 (bottom panel), 
where the interfacial modes $_li_2$ have large amplitudes
at the interface between the solid crust and the fluid core.
For the model NS05T7, magnetic modes, expediently labelled $m_k$, are also plotted.
The frequency $\bar\omega$ shown is the local one, and 
$\nu_\infty=\omega\left(1-2GM/Rc^2\right)^{1/2}/2\pi$ is also dsiplayed on the right axis
of each of the panels.}
\end{figure}

\begin{figure}
\centering
\psfig{file=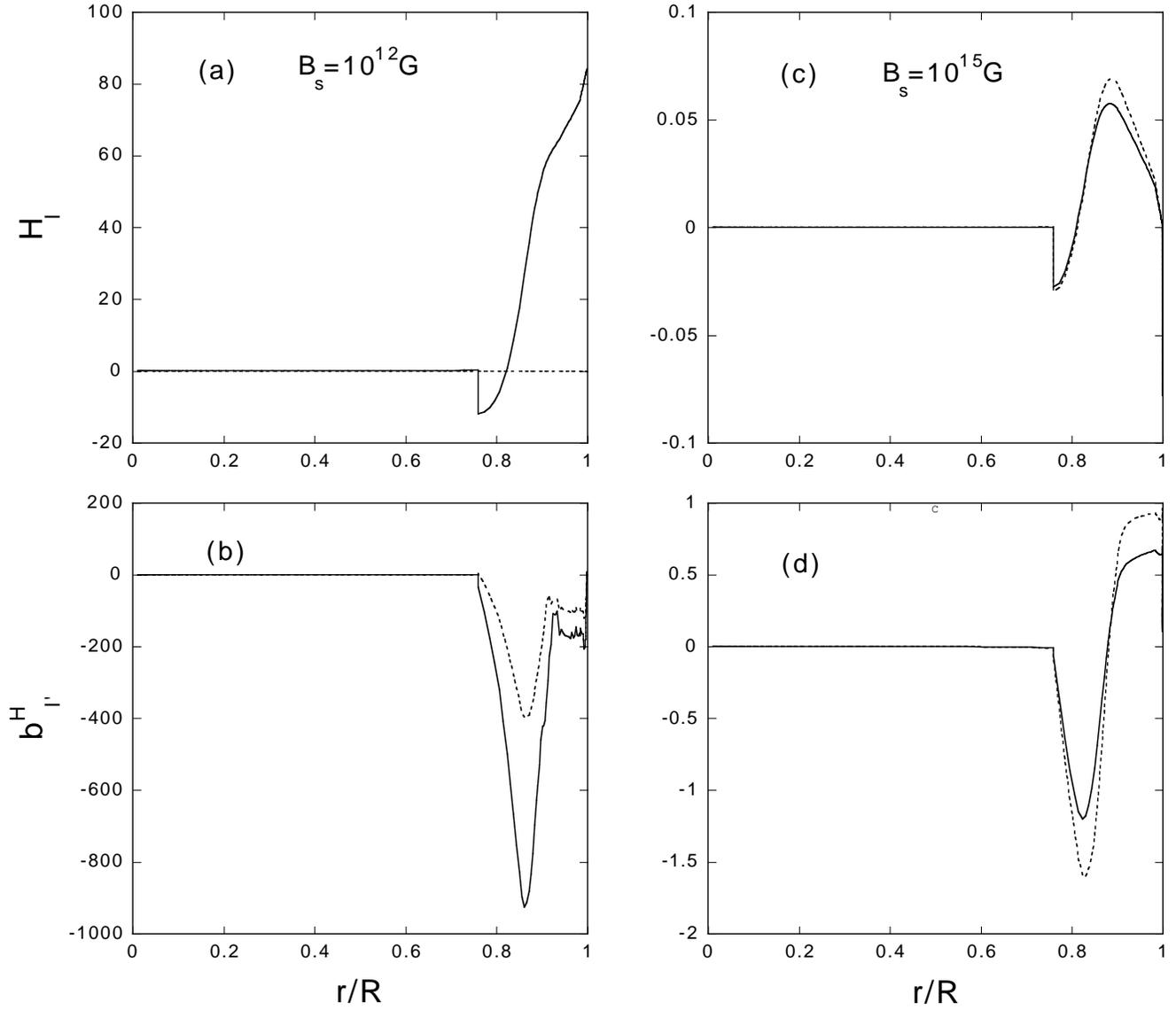,width=\textwidth}
\caption{ Eigenfunctions $H_{l}$ and $b^H_{l^\prime}$ of the axisymmetric ($m=0$)
spheroidal shear mode $_2s_1$ for $B_S=10^{12}$G in panels (a) and (b) and for $B_S=10^{15}$G
in panels (c) and (d), where
the solid lines stand for $H_{l=2}$ and $b^H_{l^\prime=1}$, and
the dashed lines for $H_{l=4}$ and $b^H_{l^\prime=3}$, respectively.
The amplitude normalization is given by $S_{l=0}=1$ at the surface.
}
\end{figure}

\begin{figure}
\centering
\psfig{file=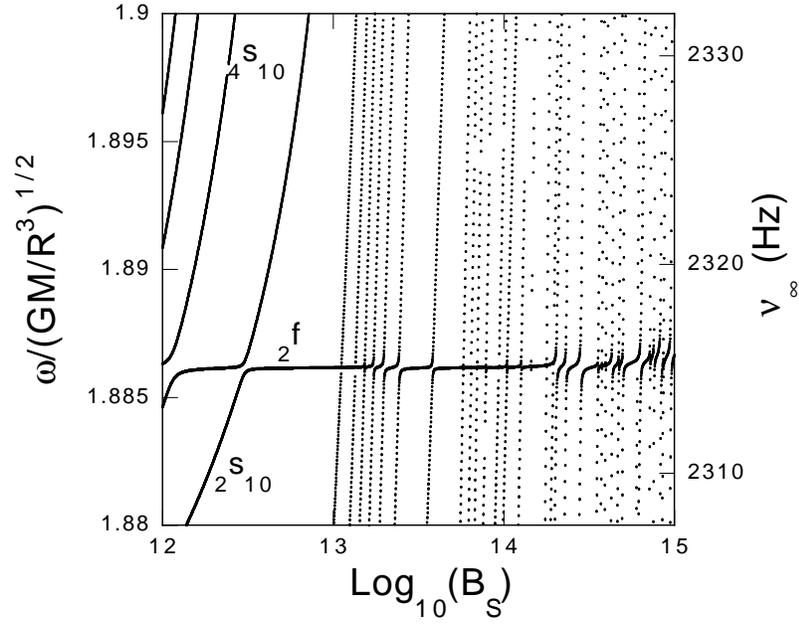,width=0.6\textwidth}
\caption{$\bar\omega$ versus $\log(B_S)$ for axisymmetric ($m=0$) fundamental $_{l=2}f$ mode
of the neutron star model NS05T7.
The $_{l=2}f$ mode suffers mode crossings with high radial order shear modes $_ls_n$
as $B_S$ increases.
The frequency $\bar\omega$ shown is the local one, and  
$\nu_\infty=\omega\left(1-2GM/Rc^2\right)^{1/2}/2\pi$ is also dsiplayed on the right axis
of the panel.}
\end{figure}

\begin{figure}
\centering
\psfig{file=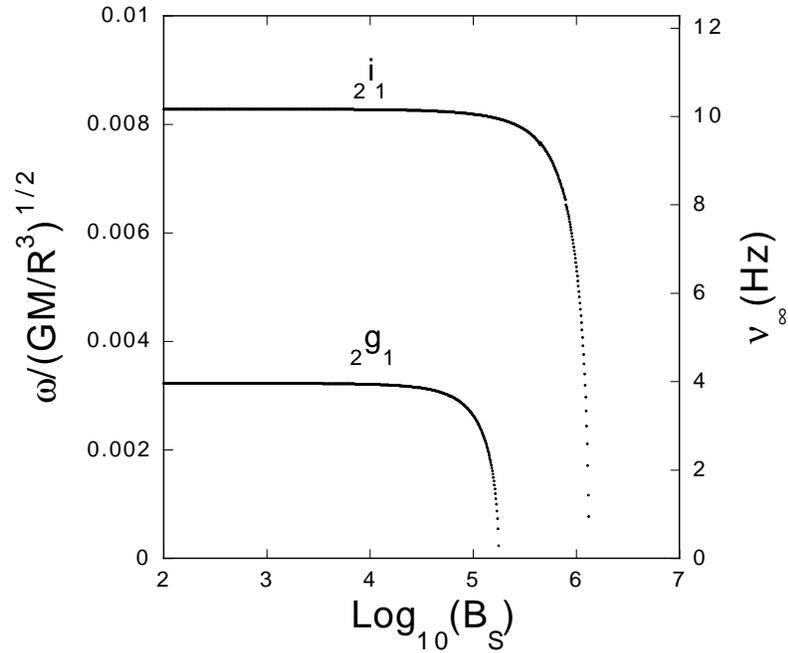,width=0.6\textwidth}
\caption{$\bar\omega$ versus $\log(B_s)$ for the axisymmetric ($m=0$)  $_{l=2}g_{n=1}$ mode
and the crust/ocean interfacial mode $_{l=2}i_1$ whose amplitudes peak at the crust/ocean interface
for the neutron star model NS05T7.
}
\end{figure}

\begin{figure}
\centering
\psfig{file=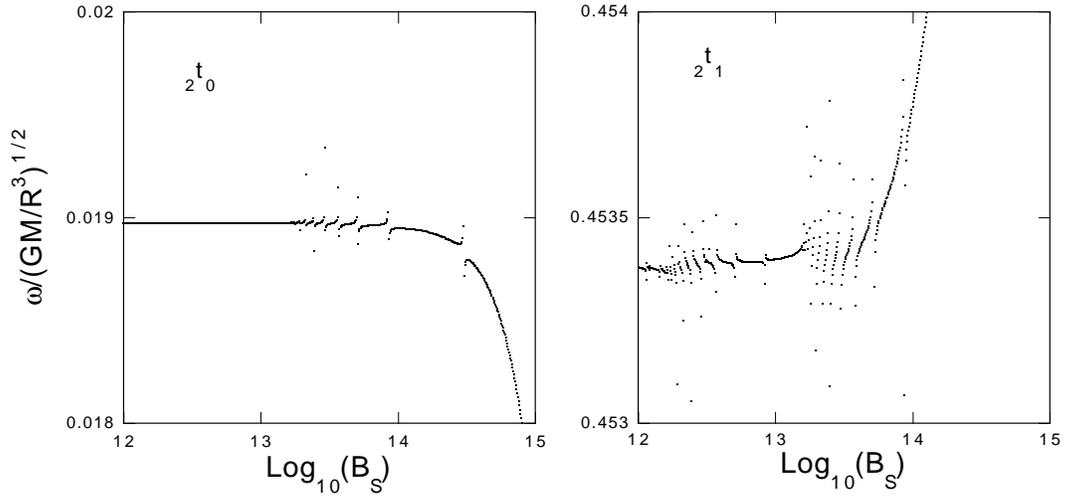,width=0.8\textwidth}
\caption{$\bar\omega$ versus $\log(B_s)$ for the fundamental torsional
mode $_{2}t_0$ (left panel) and the first overtone torsional mode $_{2}t_1$ (right panel)
of the model NS13T8.
The panels are magnifications of the corresponding parts from Figure 1.
}
\end{figure}

\begin{figure}
\centering
\psfig{file=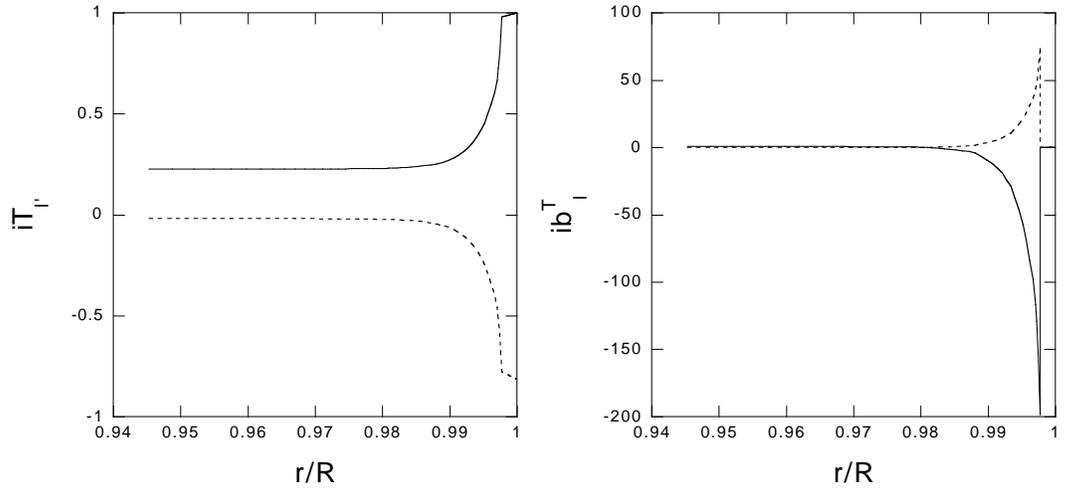,width=0.8\textwidth}
\caption{Eigenfunctions $iT_{l^\prime}$ and $ib^T_{l}$ of the $m=0$
toroidal magnetic mode of $\bar\omega=0.01557$ 
at $B_S=10^{14.48}$G for the model NS13T8, where
the solid lines stand for $iT_{l^\prime=2}$ and $ib^T_{l=1}$, and
the dashed lines for $iT_{l^\prime=4}$ and $ib^T_{l=3}$, respectively.
The amplitude normalization is given by $iT_{l^\prime=2}=1$ at the surface.
}
\end{figure}

\begin{figure}
\centering
\psfig{file=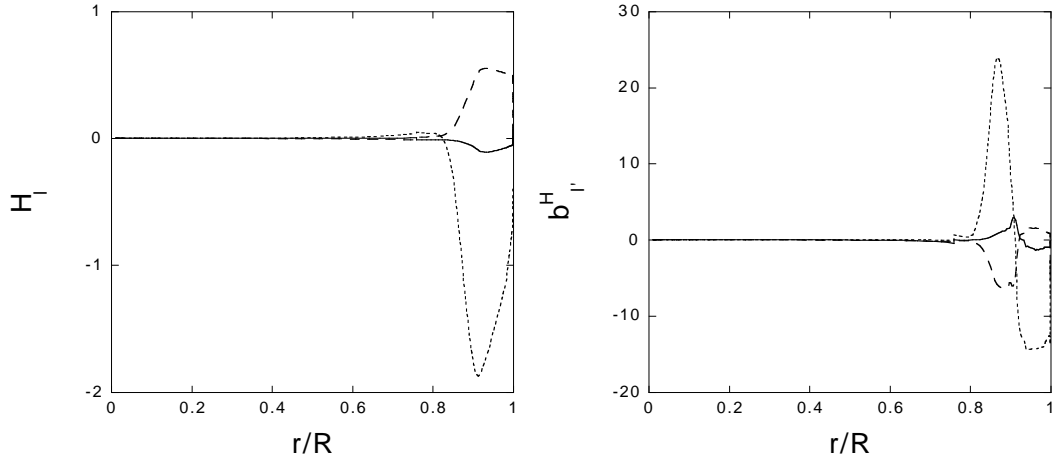,width=0.8\textwidth}
\caption{Eigenfunctions $H_{l}$ and $b^H_{l^\prime}$ of the $m=0$
spheroidal magnetic mode labeled $m_1$ at $B_S=10^{14.45}$G for the model NS05T7, where
the solid lines stand for $H_{l=2}$ and $b^H_{l^\prime=1}$, 
the dashed lines for $H_{l=4}$ and $b^H_{l^\prime=3}$, and the
dotted lines for $H_{l=6}$ and $b^H_{l^\prime=5}$, respectively.
The amplitude normalization is given by $S_{l=0}=1$ at the surface.
}
\end{figure}

\end{document}